%% file: revised.tex
\documentclass[sigplan,10pt]{acmart}
\usepackage{listings}

\usepackage{subfig}

\usepackage{float}
\usepackage[inline]{enumitem}

\usepackage{graphicx}
\usepackage{color}
\usepackage{soul}
\usepackage{times}

\usepackage{amsmath}
\usepackage{amsfonts}

\usepackage{url}
\usepackage{setspace}
\usepackage{hyphenat}
\DeclareMathOperator*{\argmaxA}{arg\,max}

\usepackage{changes}

\definechangesauthor[name={Authors}, color={red}]{au}

\definecolor{mygreen}{rgb}{0,0.2,0}
\definecolor{mygray}{rgb}{0.5,0.5,0.5}
\definecolor{mymauve}{rgb}{0.58,0,0.82}
\definecolor{mypurple}{rgb}{0.38,0,0.32}
\definecolor{myblue}{rgb}{0.1,0,0.32}

\author{Marco Cianfriglia}
\authornote{Most of the work was performed while the author was an intern at Dividiti.}
\affiliation{%
  \institution{Department of Mathematics}
  \city{Rome Tre University}
  \state{Italy}
}
\email{cianfriglia@mat.uniroma3.it}

\author{Flavio Vella}
\affiliation{%
  \institution{Dividiti}
  \city{Cambridge}
  \state{UK}  
}
\email{flavio@dividiti.com}

\author{Cedric Nugteren}
\affiliation{%
  \institution{Tom Tom}
  \city{Amsterdam}
  \state{Netherlands}  
}
\email{mail@cedricnugteren.nl}

\author{Anton Lokhmotov}
\affiliation{%
  \institution{Dividiti}
  \city{Cambridge}
  \state{UK}  
}
\email{anton@dividiti.com}

\author{Grigori Fursin}
\affiliation{%
  \institution{Dividiti and cTuning foundation}
  \city{Paris}
  \state{France}  
}
\email{grigori.fursin@ctuning.org}

\title{A model-driven approach for a new generation of adaptive libraries}

\begin{document}

\begin{abstract}
Efficient high-performance libraries often expose multiple tunable parameters to provide highly optimized routines. 
These can range from simple loop unroll factors or vector sizes all the way to algorithmic changes, given that some implementations can be more suitable for certain devices by exploiting hardware characteristics such as local memories and vector units. 
Traditionally, such parameters and algorithmic choices are tuned and then hard-coded for a specific architecture and for certain characteristics of the inputs.
However, emerging applications are often data-driven, thus traditional approaches are not effective across the wide range of inputs and architectures used in practice. 
In this paper we present a new adaptive framework for data-driven applications which uses a predictive model to select the optimal algorithmic parameters by training with synthetic and real datasets. We demonstrate the effectiveness on a BLAS library and specifically on its matrix multiplication routine. We present experimental results for two GPU architectures, and show significant performance gains of up to 3x (on a high-end NVIDIA Pascal GPU) and 2.5x (on an embedded ARM Mali GPU) when compared to a traditionally optimized library.
\end{abstract}
\maketitle
\section{Motivation}
\label{sec:intro}
Scientific HPC applications are built around monolithic parallel routines that are often customized for a specific target architecture. 
With the advent of big-data and data-driven applications such as deep learning, graph analytics or image recognition, the traditional library design looses performance portability mainly due to the unpredictable size and structure of the data. 
For example, in graph processing, the computation is dictated by vertices and edges (entities and relations); therefore, it might be hard to identify an optimal parallel strategy (i.e., data-thread mapping or partitioning) a priori~\cite{lumsdaine2007challenges,bernaschi2016scalable}.
Matrix multiplication represents another notable example where it is quite hard to determine the specific optimizations required for given input dimensions. 
Due to the ubiquity of matrix multiplications in many scientific applications, Basic Linear Algebra Subprograms (BLAS) and, in particular, the general matrix multiplication (GEMM) routine are the main target of optimizations. Several BLAS implementations provide fast performance on a target architecture by assuming a fixed data size or structure (i.e., square matrices)~\cite{Whaley00automatedempirical, nvidia2008cublas, intelMKL}. However, the matrices involved in the training of deep neural networks, for example, expose different sizes and usually rectangular shapes~\cite{deepbench}. As a consequence, it is hard to find a good optimization which takes into account the wide range of data sizes involved. In practice, most BLAS libraries often provide several GEMM implementations for specific input characteristics. Such user-transparent implementations are selected by naive heuristics based on customized decision rules. However, such solutions suffer from \textit{over-fitting} and poor performance on average.

With the wide variety of parallel architectures available on the market ranging from traditional parallel processors to accelerators (GPUs, FPGAs) and system on chips (SoCs), 
several standards have been established to enable portability for heterogeneous architectures such as OpenCL~\cite{stone2010opencl} and OpenACC~\cite{wienke2012openacc}.
However, developing generic and performance portable code has become extremely challenging, especially from an algorithmic point of view. 
Here, parametric implementations and auto-tuning techniques have partially mitigated the performance portability problem by adapting the underlying memory hierarchies and/or data thread mapping to a specific architecture. 
Within this context, a plethora of hardware-oblivious solutions have been developed~\cite{CLBlast2017, Tillet:2012:AOC:2338816.2338820, heimel2013hardware, goyal2010founding}.

This paper aims to offer a new prospective on adaptive libraries and performance portability to start addressing the problem of data-aware and architecture-aware software. 
Focusing on GPU architectures and the GEMM routine as a use case, we present a new framework based on a predictive model to select the optimal algorithm and tuning parameters to improve performance of data-driven applications. 

The contributions of the paper are summarized as follows:
\begin{itemize}
  \item we adopt a machine learning based methodology to design adaptive libraries to achieve performance portability across different datasets and hardware;
  \item we analyze several configurations of \emph{decision trees}, one of the simplest univariate supervised classifiers. This is used to select an optimized implementation by predicting the algorithm and tuning parameters;
   \item we describe three different approaches to generate training dataset to learn predictive models;
  \item we validate our study by providing exhaustive experimental results where we also evaluate the performance of the predictive models in terms of the accuracy and run-time overhead;
  \item we integrate our solution in an OpenCL BLAS library, CLBlast~\cite{CLBlast2017}, resulting in speed-ups of up to 3x and 2.5x for a high-end NVIDIA GPU architecture and an embedded ARM Mali GPU respectively;
\end{itemize}
The remainder of this paper is organized as follows. Section~\ref{sec:background} provides the background. Section~\ref{sec:mlt} describes our methodology and framework.
Section~\ref{sec:gemm-use-case} considers GEMM as a use case. Section~\ref{sec:exp} presents our exhaustive experimental evaluation. Section \ref{sec:related} discusses related work. Finally, Section \ref{sec:conclusion} summarizes the contributions of this work and outlines its future directions.

\section{Background}
\label{sec:background}
In this section, we provide the notation and basic concepts used in the paper. 
We describe the fundamentals of the decision trees classifier~\ref{sub:dt-classifier}, the generic matrix-matrix multiplication (GEMM) routine~\ref{sub:gemm} and CLBlast library~\ref{sub:clblast}. 
\subsection{Decision Tree Classifier}
\label{sub:dt-classifier}
Decision trees is a non-parametric supervised machine learning method used for classification and regression~\cite{safavian1991survey, han2011data}.
The aim is to create a model that predicts the value of a target variable by learning simple decision rules inferred from the data features.

Decision trees have several advantages:
\begin{itemize}
\item most operations on a decision tree are logarithmic in the number of data points used to train the tree;
\item they follow a ``white box model'' which is simple to understand and to interpret (unlike for example a neural network model which is more difficult to interpret);
\item models can be easily translated as if-then-else statements. 
\end{itemize}
Decision trees also exhibit some disadvantages:
\begin{itemize}
\item a decision tree might create over-complex tree that do not properly generalize the data (over-fitting);
\item small variations in the data might result in a completely different tree being generated (data perturbation);
\item from a complexity point of view, the problem of learning an optimal decision tree is known to be NP-complete. Consequently, practical algorithms cannot guarantee to return the globally optimal decision tree (low accuracy).
\end{itemize}

We use the scikit-learn library to build and analyze decision trees~\cite{Pedregosa:2011:SML:1953048.2078195}. 
This library provides several parameters in order to define different split criteria (e.g. the maximum height of tree), the minimum number of samples required to split of an internal node, and other metrics (e.g., Gini impurity) on top of an optimized version of the CART algorithm~\cite{breiman1984classification}.

\subsection{Generic Matrix Multiplication}
\label{sub:gemm}
Matrix-multiplication is one of the key components of traditional scientific applications, but also of deep learning and other machine learning algorithms.
\begin{equation}
 C = \alpha \cdot A \cdot B + \beta \cdot C \quad s.t.  \quad A \in \mathbb{C}^{M\text{x}K}, B \in \mathbb{C}^{K\text{x}N}, C \in \mathbb{C}^{M\text{x}N}
\end{equation}
where $A$ and $B$ are the input matrices, $C$ is the output and $\alpha$ and $\beta$ are constants. The operands $A$ and $B$ can be optionally transposed.  
In general, a matrix multiplication is represented in terms of size by the tuple $(M, N, K)$ describing the sizes of the matrices involved. The complexity is $\mathcal{O}(M \cdot N \cdot K)$~\cite{grama2003introduction}.
A naive algorithm sequentially calculates each element of $C$ by using three nested loops. However, in practice, fast computation can be achieved by maximizing data-reuse. In general, parameters, such as tiling, threads organization and scheduling can influence the performance~\cite{6354699}. For example, for a specific target architecture different values of tile sizes strongly impact data-reuse in local memories. Tuners explore huge search space of such parameters in order to find the best performance for a specific input size and architecture. Notable solutions and techniques about BLAS libraries and auto-tuning are reported in Section~\ref{sec:related}.

\subsection{CLBlast Library}
\label{sub:clblast}
CLBlast is a modern, lightweight, fast and tunable OpenCL BLAS library written in C++11~\cite{CLBlast2017}. It is designed to leverage the full performance potential of a wide variety of OpenCL devices from different vendors, including desktop and laptop GPUs, embedded GPUs, and other accelerators. 
The library implements BLAS routines: basic linear algebra subprograms operating on vectors and matrices. Specifically to GEMM, CLBlast provides two kernels: a ``direct'' kernel covering all GEMM use-cases, and an ``indirect'' kernel making several assumptions about the layout and sizes of the matrices. The ``indirect'' kernel cannot be used on its own and requires several helper kernels to pad and/or transpose matrices to meet these assumptions. Thus, there is a performance trade-off between running the more generic ``direct'' kernel versus the specialized  ``indirect'' kernel ($\mathcal{O}(n^3)$) plus several helper kernels ($\mathcal{O}(n^2)$).
\begin{figure}[!ht]
  \centering
  \includegraphics[width=1.06\columnwidth]{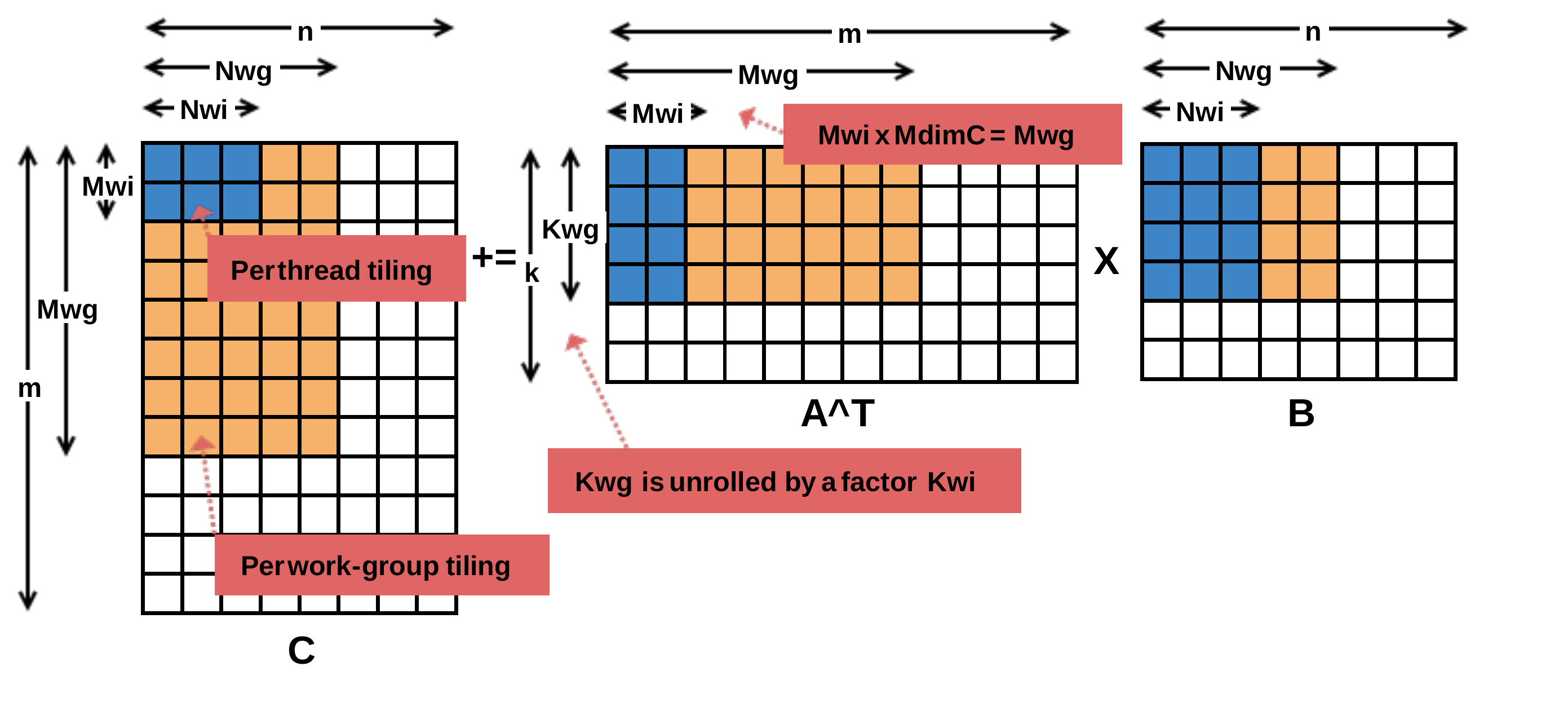}
  \vspace{-20pt}
  \caption{GEMM in CLBlast. The blue area indicates work done by a single thread, the orange area indicates work done per OpenCL work-group. Image taken from~\cite{CLBlast2017}.}
  \label{fig:gemm_clblast}
\end{figure}
Furthermore, at a kernel level, there are many tunable parameters, 6 of which are illustrated in Figure~\ref{fig:gemm_clblast}. The parameters define for example the work-group sizes in 2 dimensions ($M_{wg}, N_{wg}$), 2D register tiling ($M_{wi}, N_{wi}$), vector widths of both inputs, loop unroll factors ($K_{wi}$), and how to use the local memories and caches. In total the search space for a GEMM kernel can easily grow to a hundred thousand realistic combinations. For more details we refer to the CLBlast and CLTune papers~\cite{CLBlast2017, CLTune2015}.

\section{Methodology and framework}
\label{sec:mlt}
In this section, we introduce a methodology and framework for generating a model-driven optimisation for input-aware adaptive libraries. 
The idea is to learn a model based on the inputs characteristics of a specific problem. 
For this purpose, we identify three desirable characteristics of the framework.

First, we should be able to select the best solution (algorithm and/or implementation and/or configuration) among multiple possible choices according to an objective function.
Formally, let $A$ be a finite set of solutions of a particular problem (e.g. matrix multiplication, graph traversal). 
Let $f_{a}: I \rightarrow \mathbb{R}$ be an objective function (e.g. floating point operations per second (FLOPS) or traversed edges per second (TEPS)), where $I$ is  multidimensional input domain for $A$. 
For example, $I$ can represent the set of all triples $(M, N, K)$ which describe the GEMM operands.
The goal is maximizing $\overline{a} = \argmaxA_a  f_a(i)$ for each $i \in I$.

Second, we should be able to build a predictive model starting from the training dataset $\overline{A}$ consisting of all, or representative, optimal solutions $\overline{a}$.

Third, we should be able to generate code implementing the model.
Furthermore, the generated implementation should satisfy the following requirements:
\begin{enumerate}
 \item \textbf{correctness and soundness}: the model should be able to manage the same input domain of the original library;
 \item \textbf{cost-effectiveness}: the generated code should have negligible overhead. In fact, the cost of selecting the best routine must be lower than the improvement. Formally, $f_{\overline{a}}(i) + c_{\overline{a}} < f_a(i)$ where $c_{\overline{a}}$ is the cost to select $\overline{a}$.
\end{enumerate}

\subsubsection*{Framework design and workflow}
Logically, the framework is composed of two separate phases:
\begin{enumerate}
 \item during the off-line phase, we create a training dataset, build a predictive model from this dataset and integrate the model into the target library;
 \item during the on-line phase, we use the learned model integrated into the library.
\end{enumerate}
Decoupling the computationally expensive off-line phase from the on-line phase means that we can use different training datasets, as well as machine learning techniques for building models. Also, there is no need to package the machine learning framework with the target library. In our implementation, for example, a decision tree is represented by a complex if-then-else statement.

\begin{figure}[ht]
  \centering
  \includegraphics[width=1.06\columnwidth]{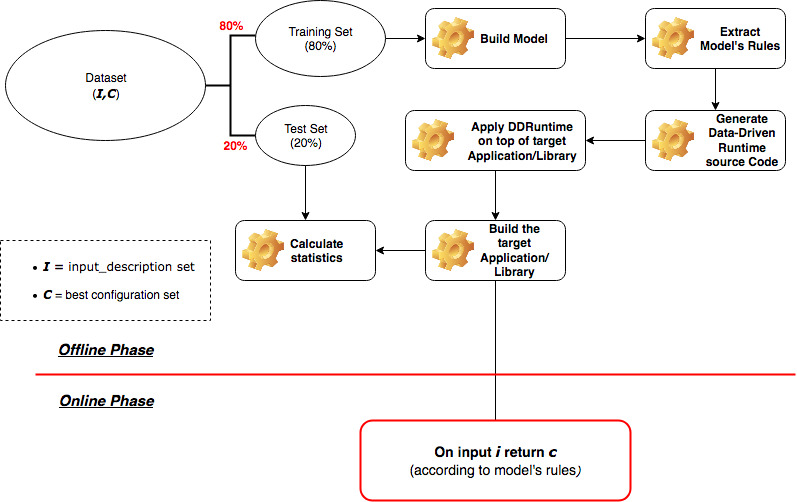}
  \vspace{-10pt}
  \label{fig:DDRuntime}
  \caption{An overview of the proposed framework, showing the separation between the off-line (training) and on-line (deployment) phases.}
\end{figure}

\subsubsection*{Datasets}
\label{sec:collect_dataset_mlt}
We define a dataset $D$ as a collection of pairs $(I,C)$ where $I$ is the input description and $C$ is the corresponding class description.
The input description $I$ contains information about the size (e.g. the triple $(M,N,K)$ for GEMM), the structure (e.g. the density), and any additional information or metrics that can characterize the input (e.g. the data layout).
The class description $C$ represents the best algorithm/implementation/configuration for a given input according to the objective function.
Roughly speaking, the dataset is a collection of benchmarking results over a specific set of input characteristics and a given metrics. 
For example, for GEMM, the metrics are usually FLOPS or FLOPS per Watt, while $C$ is simply the best implementation/configuration for the given metrics.

Several strategies can be used for generating the dataset:
\begin{enumerate}
 \item \textbf{synthetic}: $I$ is generated according to a specific rule;
 \item \textbf{real-world}: $I$ is collected from real workloads;
 \item \textbf{hybrid}: $I$ is a mix of synthetic and real-world instances. 
\end{enumerate}

For example, for GEMM, a synthetic dataset can be generated by processing all triples $(M, N, K)$ for which $M$, $N$, and $K$ are all the powers of two within a domain; a real-world dataset can be collected by profiling the operands involved in a specific application such as a deep neural network (e.g. see DeepBench~\cite{deepbench}).
For graph traversal problems, a synthetic dataset can be generated from R-MAT graphs~\cite{chakrabarti2004r}, while a real-world dataset can be collected from graph applications (see the SNAP dataset~\cite{snapnets}).

A dataset $D$ is usually divided into two disjoint subsets $X$ and $Y$ such that $D = \lbrace X \rbrace \cup \lbrace Y \rbrace$ via random sampling. 
The subsets $X$ and $Y$, namely the \textsl{training dataset} and the \textsl{test dataset}, can contain for example $80\%$ and $20\%$ of $D$ respectively~\cite{han2011data}.

The quality of the dataset plays a central role in the learning phase and strongly depends on the real use-case (see the next section).

\subsubsection*{Model and code generation}
Several models can be created from the same training dataset $X$ and evaluated over the test dataset $Y$. 
To learn a model, we identify the set of \emph{features} and \emph{labels} (or \emph{classes}) in the training dataset; we select the input descriptions $I$ as features and the configuration descriptions $C$ as labels. 
Then, the model is learned according to the specific machine learning framework and algorithm used. 
Specifically, in our implementation, we use the CART algorithm to build a decision tree, but this can be replaced with any other suitable technique according to the problem at hand.
Traditional machine learning techniques, such as cross validation, can also be applied in this phase.  
In the case of a simple decision tree, the system automatically extracts all the rules defined in internal nodes as well as the configurations represented by the leaves of tree. 
From the learned model, this procedure generates source code in the form of an if-then-else tree, which then gets automatically integrated into the target library.

\section{A model-driven adaptation for GEMM} 
\label{sec:gemm-use-case}

We present a \emph{proof of concept} to show the effectiveness of our methodology applied to a case study. 
We investigate parallel matrix multiplication since it is ubiquitous in several HPC applications ranging from computational science (e.g., fluid dynamics) to deep learning and graph analytics. 
\subsection{Dataset}
First, we define the dataset class description $C$ according to the target library capabilities. CLBlast, in combination with CLTune, provides multiple algorithmic choices defined by tuning parameters. 
Table~\ref{tab:clblast_stats} summarizes CLBlast characteristics for \textit{xgemm} and \textit{xgemm direct} routines. 
According to these characteristics, the number of possible different classes for each triple $(M,N,K)$, which corresponds to an entry in $I$, is bounded by $\sum_{j=0}^{|A|} \hat{A}_j$ where $\hat{A}_j$ is the set of the legal assignments within of the search space of the $j^{th}$ algorithm. This distinction is necessary because some parameter combinations are invalid for a specific input or architecture.
For example, a target architecture may not support too large an OpenCL work-group size or have limited local memory available.

Actually, the number of classes can be extended by increasing the search space of tunable parameters. Note that extending the search space bounds may require managing possible illegal parameters which might violate the correctness and soundness rule: each class in the dataset must be a valid configuration for each entry in $I$. For GEMM in CLBlast, we do not have to manage the problem of finding the best configuration ourselves. Instead, we use the existing exhaustive approach provided by the CLTune tuner~\cite{CLTune2015} to find the best configurations for its two GEMM kernels (\textit{xgemm} and \textit{xgemm direct}) measured in terms of FLOPS.
Each entry in $D$ is a pair (($M,N,K$), $\overline{a}$) where $\overline{a}$ is the best kernel represented by its tunable parameter configuration for the given $(M,N,K)$. From the CLBlast point of view, this means applying the tuner for the two GEMM kernels for a given $(M,N,K)$ and recording the best solution among them. This approach is expensive when the size of $I$ becomes significant. It is possible to trade-off quality versus time by sampling randomly from the set of tuning parameters. In this paper, however, we explore the entire search space in order to simplify the analysis during the generation of the model by avoiding perturbations on the models due to random sampling. This allows us to provide a fairer comparison among different datasets and generative model strategies.

\begin{table}[!ht]
  \centering
  \small
  \begin{tabular}{|l|c|c|}
    \hline
    \textbf{Kernels} & \textbf{Tunable Parameters} & \textbf{Search Space Size} \\ \hline
    Gemm             & $14$                          & $8748$                       \\ \hline
    Gemm direct      & $9$                           & $3888$                       \\ \hline
  \end{tabular}
  \caption{Tuning size statistics as used for this case-study.}
  \label{tab:clblast_stats}
\end{table}

Second, we determine the input descriptions $I$ of the triples $(M,N,K)$,  and, consequently, the size and other characteristics of $D$.
We provide one real-world dataset, and two different strategies for generating synthetic ones.

For the real-world dataset (friendly named \textit{AntonNet}), we gather the sizes of the GEMM operands involved in popular deep neural networks: AlexNet~\cite{krizhevsky2012imagenet}, GoogLeNet~\cite{szegedy2015going} and SqueezeNet~\cite{iandola2016squeezenet}. Specifically, we collect the sizes for the batch sizes ranging from 2 to 128 with a step of 2. 
This dataset consists of roughly $460$ different triples, with $35$\% of them having $K=1$. The other shapes are mostly rectangular.

We also generate synthetic datasets to be able to learn more generic models. In our experiments, we use two strategies that differ in terms of the distance between dataset points $(M,N,K)$, viewed as 3D coordinates in the Euclidean space:
\begin{enumerate}
 \item \emph{grid of two} (\emph{go2}): composed by $(M,N,K)$ triples where the values range from 256 to 3840 with a step of 256. This dataset is approximately 8 times larger than \emph{AntonNet}. 
 \item \emph{power of two} (\emph{po2}): composed by $(M,N,K)$ triples where the values are powers of 2 ranging from 64 to 2048.  This dataset is less dense than \emph{go2}.
\end{enumerate}

While we can easily calculate the size of each dataset $I$, the number of classes $C$ strongly depends on the architecture. For example, even if \emph{AntonNet} is smaller than \emph{go2}, the number of classes is $3$/$4$ times larger for the architectures in our study (see the first four columns of Table~\ref{table:stats-p100} and Table~\ref{table:stats-mali}). The main reason is that the matrices in \emph{AntonNet} have irregular sizes and therefore require more unique configurations than the matrices in the synthetic datasets.
The relation between the matrix sizes and the classes, as well as how to determine representative entries for a dataset, will be subject of further studies.

\subsection{Model and code generation}
A decision tree classifier usually offers multiple implementation choices in order to build a more accurate model. 
In our case, the parameters that we used for the training are $L$ and $H$. 
$L$ is the minimum number of sample per leaf required for a class to become a leaf node. This means if a class occurs one time in the dataset and $L=2$ (or higher) that class will not become a leaf in the decision tree. Scikit also allows to set up a normalized $(0,1)$ percentage over the total number of classes. 
For example with $L=0.1$ a class to be a leaf must occur in the $10\%$ of the dataset. A small values of $L$ usually means the tree will overfit, whereas a large value will build more generic trees from learning the data.

$H$ is the maximum height of the decision tree. If None, then the nodes are expanded until all the leaves are pure (all the value of the feature in the node comes from a
single class) or until all leaves contain less than $L$ samples.

To evaluate the accuracy and the performance of our approach, we trained several decision trees by tuning $L$ and $H$. 
Hereafter, we provide an experimental study for the evaluation of all the possible assignments of such parameters. 
For this case-study, we also developed a Python program to extract other features and statistics of the models that cannot be directly extrapolated from scikit-library.
Examples include number of leaves or the height of the decision tree.
The same program is also responsible of traversing the decision tree, extracting the rules defined into internal nodes, and all the configurations of the corresponding leaves.
Consequently, the program automatically generates the corresponding C++ source code which implements the trained model in the form of an if-then-else statement. 
At the end of this process, the code is compiled into the library, such as CLBlast for this case-study. 

\section{Experimental Results}
\label{sec:exp}
The experiments reported below aimed at investigating the following aspects:
\begin{enumerate}
\item the quality of the models in terms of accuracy;
\item the quality of the models in terms of the impact of misclassification;  
\item the overhead of the decision tree (if-then-else statement) generated by our framework;
\item the performance of the model-driven CLBlast library against the default version tuned for a specific matrix size.
\end{enumerate} 
We first evaluate several models generated according to the strategies described in Section \ref{sec:gemm-use-case}.
Specifically, we generate and analyze the trained models by varying the maximum height and the minimum number of samples per leaf. The possible assignments of the height $H=\lbrace 1, 2, 4, 8, Max \rbrace$, where $Max$ means that there are not restriction on the height of the tree. 
The set of the possible assignments of the minimum number of samples per leaf is $L = \lbrace1,2,4, 0.1, 0.2, 0.4, 0.5 \rbrace$. 

Concerning the comparison among CLBlast versions, we refer to \textit{peak} when we report the best performance of CLBlast tuned for a generic matrix $(M,N,K)$.
This operation requires to run the tuner for both gemm routines. 
Notice that the tuner returns the kernel time to perform the matrix multiplication only. 
In the case of \textit{xgemm}, this does not include the time required to perform auxiliary kernels, thus it represents a performance upper bound of CLBlast.
The peak of the tuner gives an estimation of how much the performance of a model is far away from the possible best. This information also reflects the ability of the code to adapt to the architecture for a given input size. 

We refer to CLBlast \textit{default} when we use the CLBlast with the optimal parameters for a default matrix size which corresponds to $M$=$N$=$K$=$1024$ for \textit{xgemm} and $M$=$N$=$K$=$256$ for \textit{xgemm direct}. 
In CLBlast, the mechanism for switching \textit{xgemm direct} and \textit{xgemm} kernel is based on a value of threshold. 
Such threshold takes into account the sizes of the operands involved in the multiplication.
This approach basically implements a linear cut of the space represented by the triples $(M,N,K)$ by assigning one gemm implementation and its own configuration.  
Finally, we refer to \textit{model}, when we report the performance of our model-driven CLBLast version.
To automatize the workflow of our framework, we used Collective Knowledge technology~\cite{7459430} for generating the datasets, learning the models and evaluating their performance.

\subsection{Hardware setup}
We focused on two different GPU architectures: a high-end NVIDIA Tesla P100 based on the Pascal architecture and an embedded ARM Mali-T860 based on the Midgard architecture. 
In Table~\ref{table:hardware}, we report a summary of the main characteristics of both architectures.
For the ARM GPU, we did not generate the \emph{go2} dataset due to the limited amount of hours available.

\input{table_hardware}

\subsection{Accuracy and Misclassification}

To estimate the quality of the models, we calculate the accuracy by using scikit-learn. 
The accuracy is a standard measure for classification problems with the aim of providing a measure of the quality of a given model in terms of right predictions on the test dataset.
It is defined as the ratio between the number of right prediction and the total number of instances in the test dataset. 
Therefore, it allows to validate and evaluate different models since the classes for each entry are known \textit{a priori}.
In our scenario, the classes are represented by the set of configurations, and implicitly by the corresponding gemm implementation as we found out through the tuner. 
For two different consecutive triples (i.e. ($256$, $256$, $256$) and ($256$, $512$, $256$)) such configurations might be likely similar to each other.
In some case, we noticed that the best configuration for a specific triple ($M$,$N$,$K$) achieves good performance for the nearest triples.
In those cases, a model likely selects a configuration $C'_{m_i} \ne C_{m_i}^{best}$ that is not too far away in terms of performance from the optimum. 
However, from classification task prospective that represents a misclassification. 
For this kind of applications, accuracy does not give a good estimation of the real performance of the model since it does not take into account the impact of the misclassification. 
To overcome this problem, we defined two metrics in order to measure the real performance of the models over the test dataset.
The first metric is defined as the average of the ratio between the performance of a model over the peak of performance of the tuner.
Likewise, the second one takes into account the performance of a model over the performance of the tuned version of CLBlast. 
We denote them as \textbf{DTPR} (`decision tree peak ratio') and \textbf{DTTR} (`decision tree tune ratio') respectively.
\textbf{DTPR} metrics provides a more accurate estimation of the models as it is able to quantify the performance of a class also in the presence of misclassification.

\subsection{Models evaluation}
\begin{figure*}[]
  \subfloat[Nvidia P100]
  {
    \includegraphics[scale=0.3]{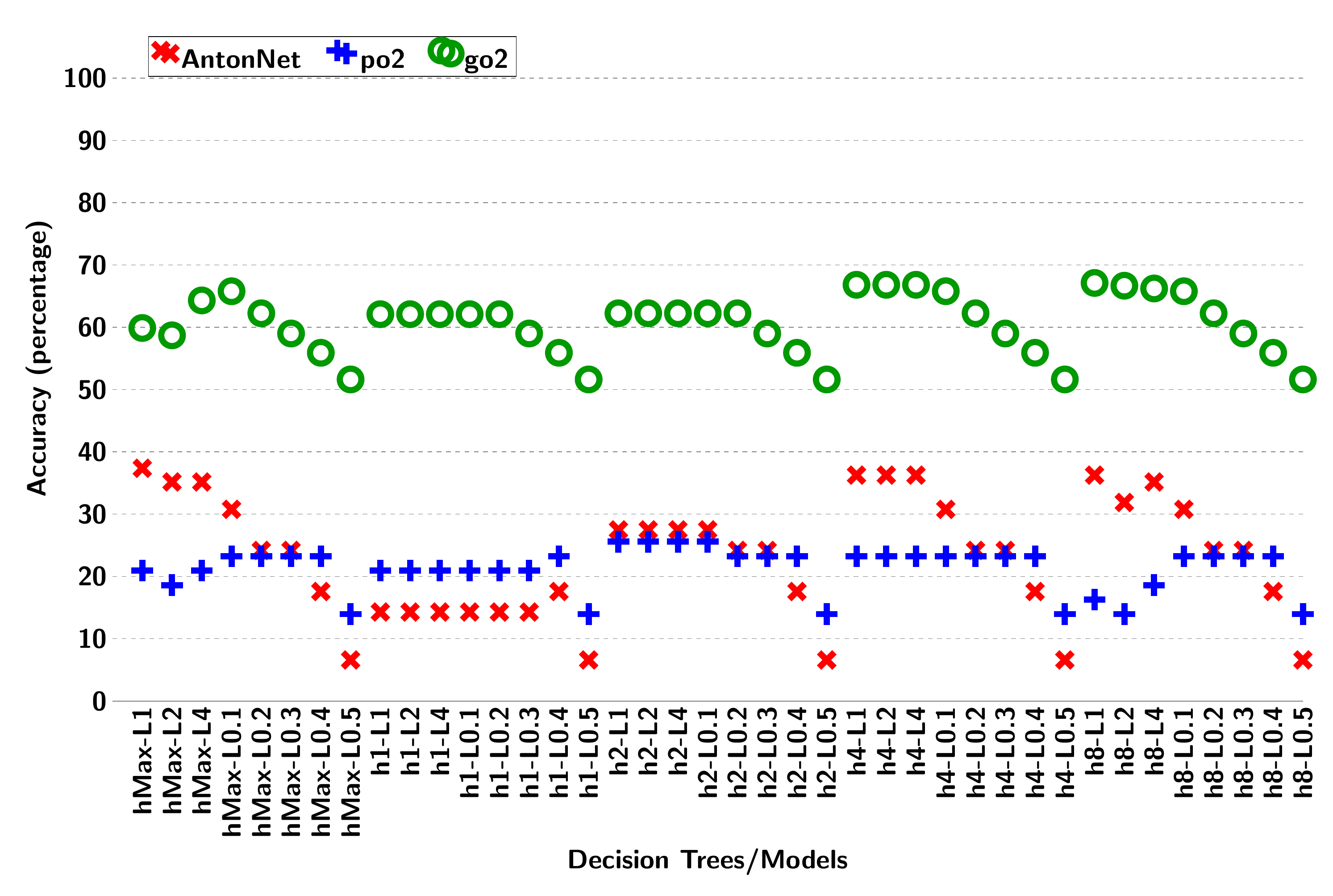}
    \label{fig:acc_p100}
  }
  \subfloat[ARM Mali-T860]
  {
    \includegraphics[scale=0.3]{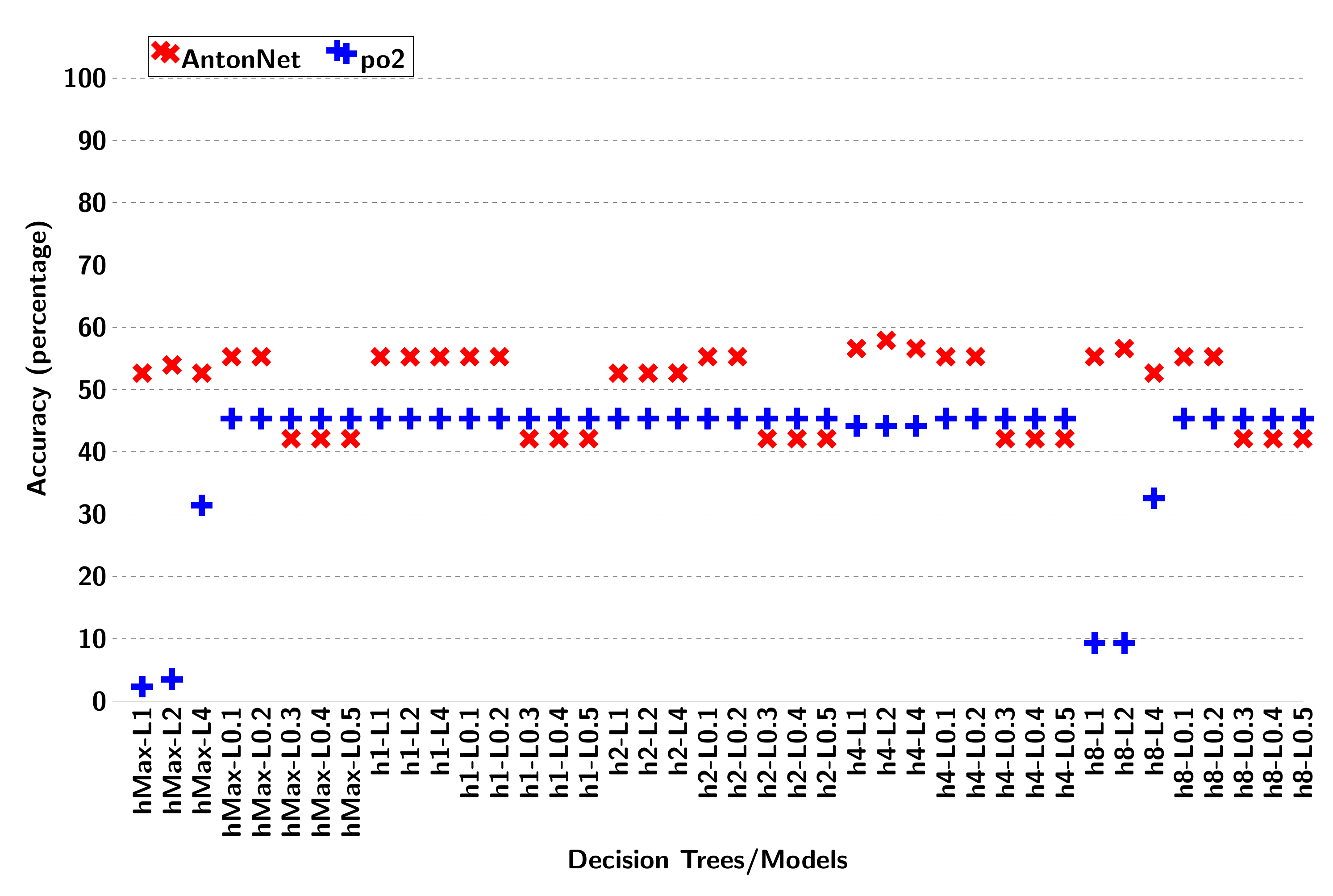}
    \label{fig:acc_mali}
  }
  \caption{Accuracy evaluation of the models generated by varying $H$ and $L$ parameters on \emph{go2} (Nvidia only), \emph{po2} and \emph{AntonNet} dataset.}
  \label{fig:accuracy}
\end{figure*}

\begin{figure*}[]
  \subfloat[Average performance ratio between the model-driven and the peak of \newline the tuner of CLBlast (\textbf{DTPR}).]
  {
   \includegraphics[scale=0.3]{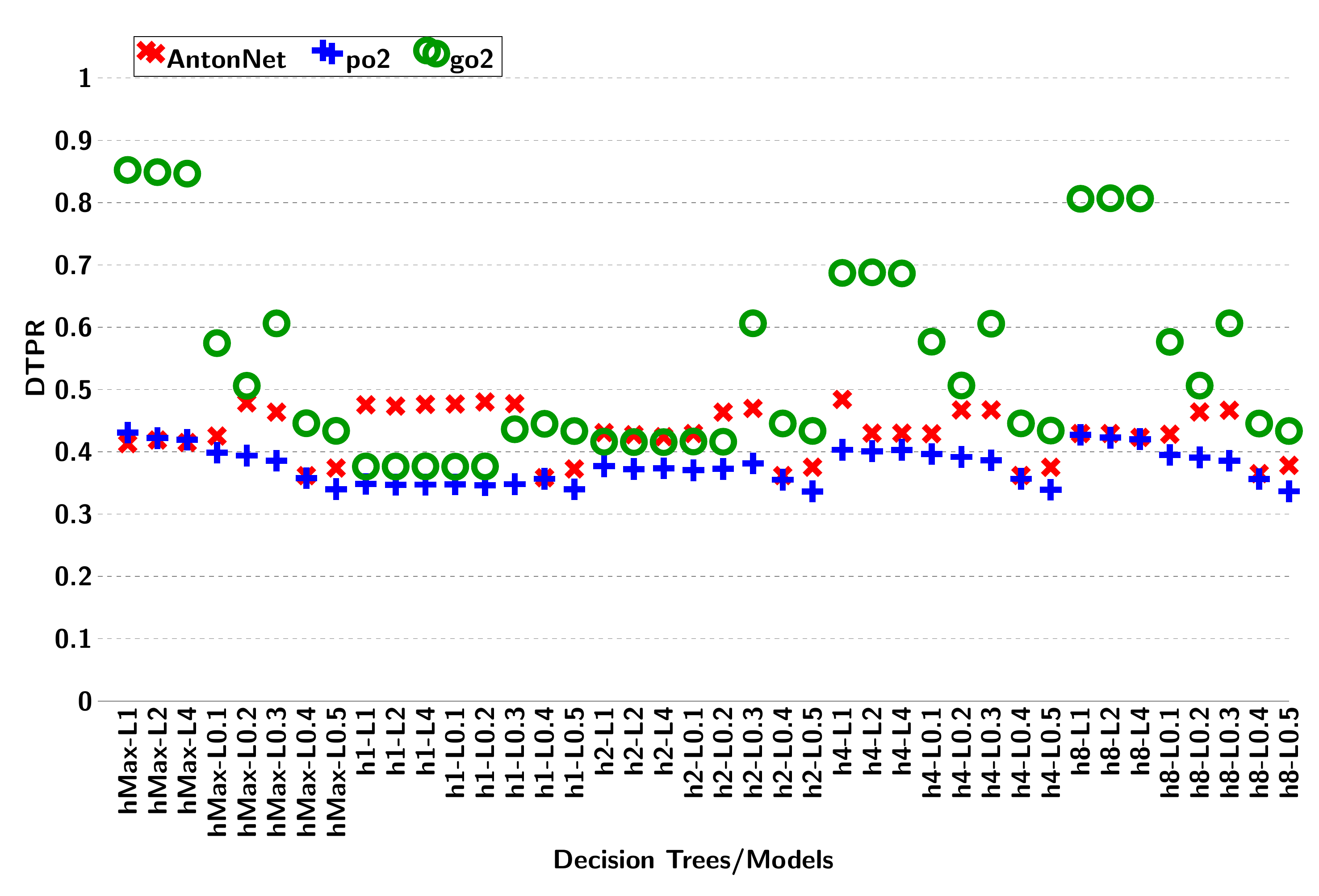}
    \label{fig:dtpr_p100}
  }
  \subfloat[Average performance ratio between the model-driven and the tuned version of CLBlast (\textbf{DTTR}).]
  {
    \includegraphics[scale=0.3]{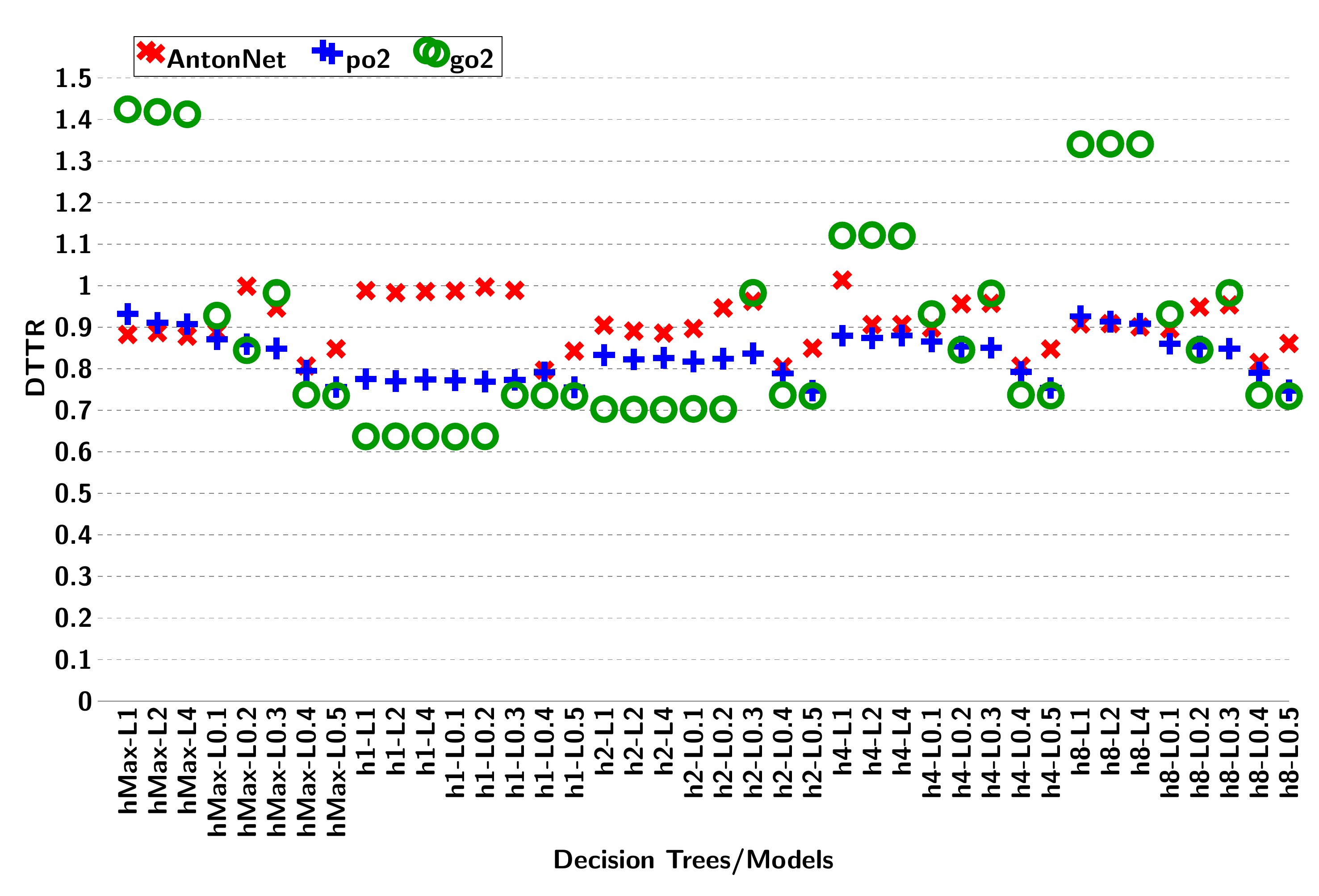}
    \label{fig:dttr_p100}
  }
  \caption{Evaluation of the impact of misclassification of the models generated by varying $H$ and $L$ parameters on \emph{go2}, \emph{po2} and \emph{AntonNet} dataset on Nvidia P100.}
  \label{fig:misclassification-nvidia}
\end{figure*}

\begin{figure*}[]
  \subfloat[Average performance ratio between the model-driven and the peak of \newline the tuner of CLBlast (\textbf{DTPR}).]
  {
   \includegraphics[scale=0.3]{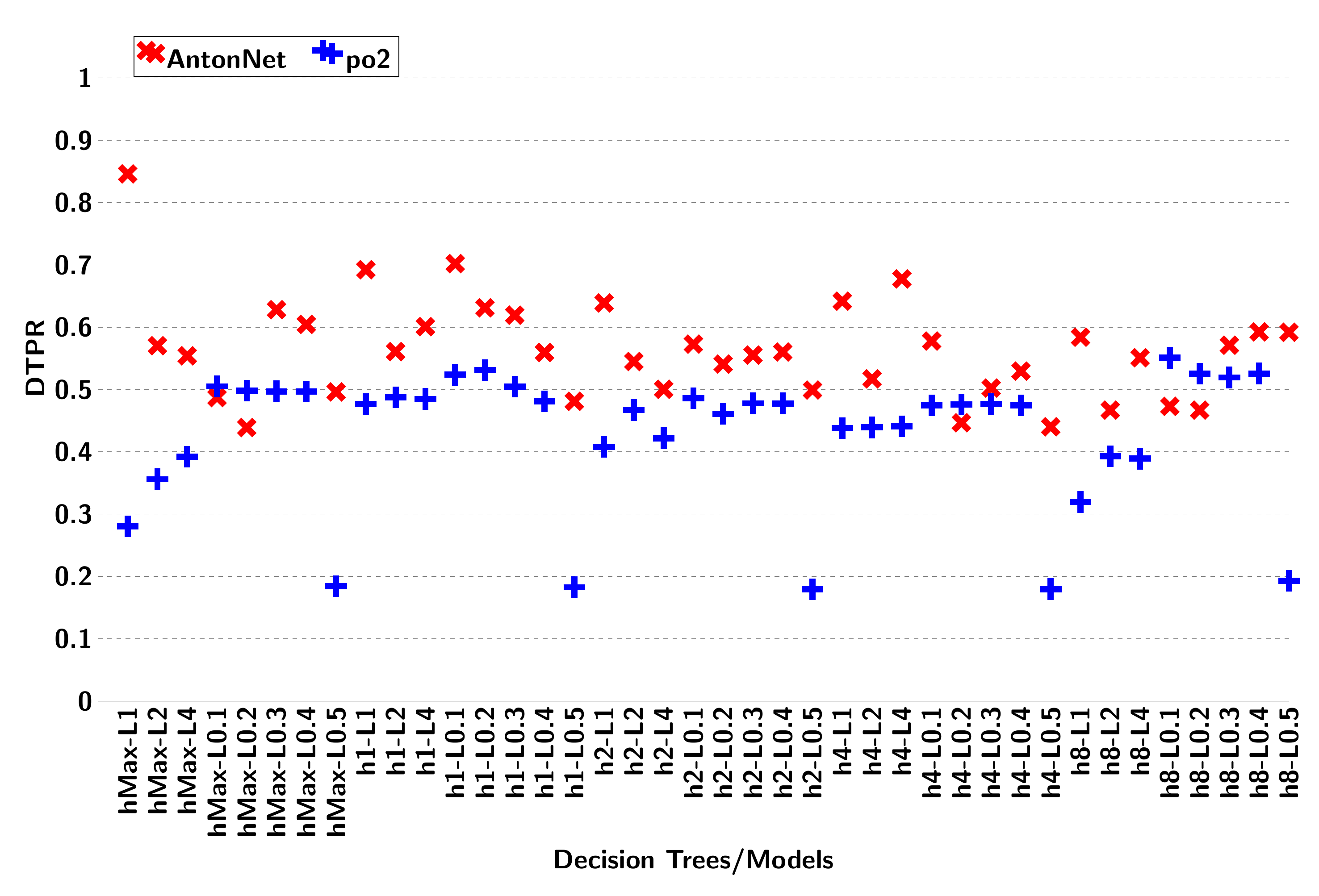}
    \label{fig:dttr_mali}
  }
  \subfloat[Average performance ratio between the model-driven and the tuned version of CLBlast (\textbf{DTTR}).]
  {
    \includegraphics[scale=0.3]{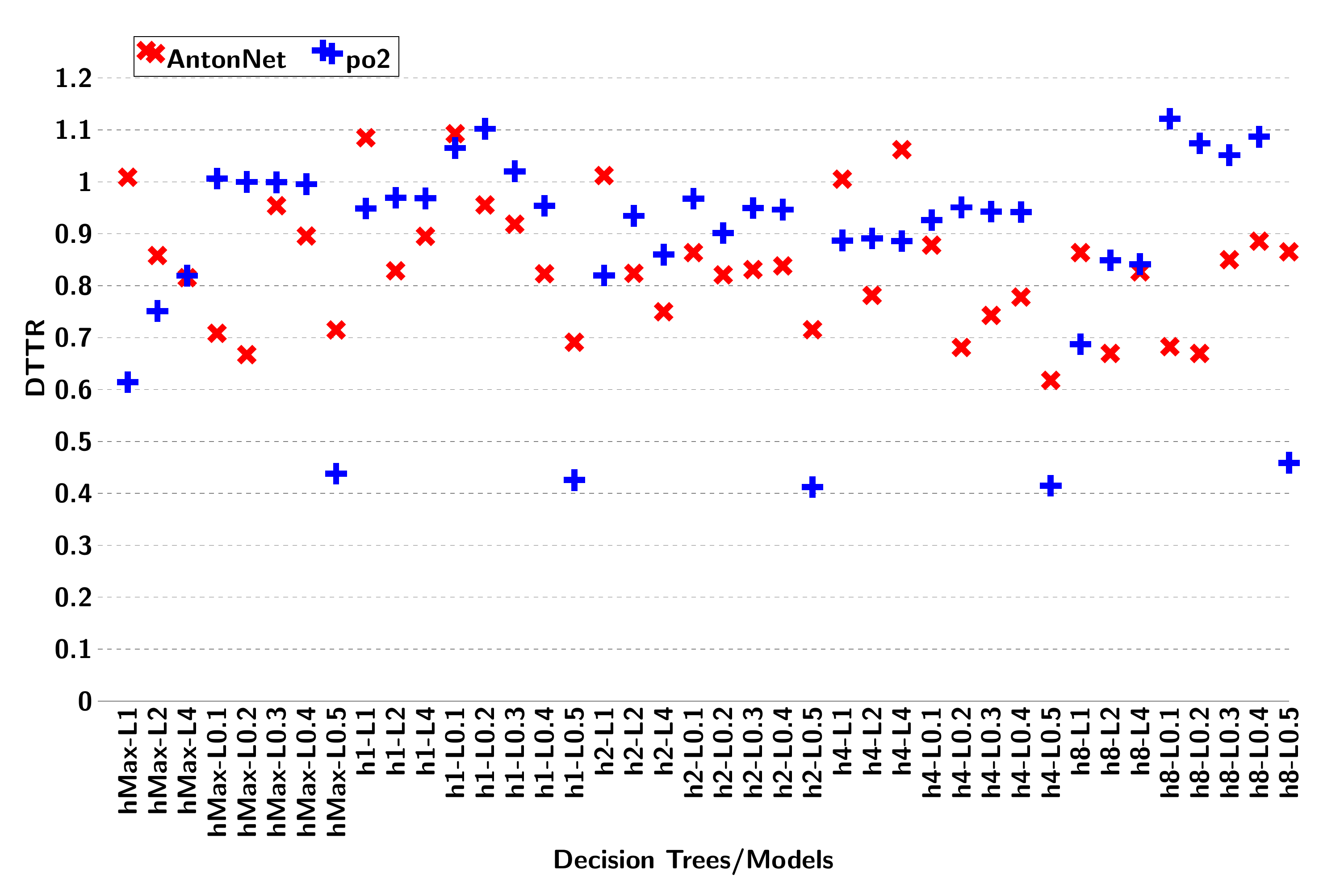}
    \label{fig:dtpr_mali}
  }
  \caption{Evaluation of the impact of misclassification of the models generated by varying $H$ and $L$ parameters on \emph{po2} and \emph{AntonNet} dataset on ARM Mali-T860.}
  \label{fig:misclassification-arm}
\end{figure*}

We start our analysis by measuring the accuracy of several models learned from our datasets by varying $H$ and $L$ parameters. 
Models should be able to predict the right class among up to $82$ differ classes (see the sum per row of the columns $3$ and $4$ in Table~\ref{table:stats-p100} and Table~\ref{table:stats-mali}).
Specifically, Figure~\ref{fig:accuracy} shows the accuracy (y-axis) of all the models (x-axis) generated by our framework for the Nvidia P100 (Figure~\ref{fig:acc_p100}) and the ARM Mali-T860 (Figure ~\ref{fig:acc_mali}).
We first noticed that a denser and regular dataset, like \emph{go2}, has a higher accuracy than a more sparse dataset like \emph{po2}. 
Unexpectedly, on the Mali GPU, \emph{AntonNet} shows a better accuracy. 
In general, we observe that the accuracy mainly depends on the distribution between gemm kernels and the number of unique configurations in the dataset 
(see Table~\ref{table:stats-p100} and Table~\ref{table:stats-mali}).
An unbalanced distribution of such configurations can be observed both \emph{AntonNet} and \emph{po2} on the Nvidia P100. 
For example, by looking at Table \ref{table:stats-p100} (columns 3-4), the configurations in these datasets mainly correspond to \textit{xgemm direct} kernel. 
The reason is that Nvidia P100 has enough resources to perform \textit{xgemm direct} in the most of the cases.
Thus, in this specific case, the classes corresponding to \textit{xgemm} will be hardly represented in the model even if the model is trained with a low value of $L$.
Contrarily, on the ARM GPU the configurations of \emph{AntonNet} are more uniformly distributed among the gemm implementations (see Table \ref{table:stats-mali}). 
From the results we observed, $H$ and $L$ parameters do not impact on the accuracy significantly. 
As an example, Figure~\ref{fig:acc_p100} shows the same trend for \emph{go2}, \emph{po2} and \emph{AntonNet} datasets even if $L$ parameter changes.
Summarizing, the model learned from \emph{go2} with $H=8$ and $L=1$ achieves the highest accuracy on the Nvidia GPU, meanwhile the model $H=4$ and $L=1$ trained from \emph{AntonNet} represents the best for the ARM GPU. 
\input{table_summary}
Accuracy experiments indirectly shows that the accuracy decreases in the presence of an increasing unbalancing distribution of the kernels and configurations.
Contrarily, \textbf{DTPR} and \textbf{DTTR} experiments indirectly provide a measure of the similarity between classes (kernel and configuration) in terms of performance: this allows  measuring the impact of the misclassification. 
Figure~\ref{fig:misclassification-nvidia} and Figure~\ref{fig:misclassification-arm} show \textbf{DTPR} and \textbf{DTTR} values for each model (see the x-axis of the figures). 
Unlike the accuracy, the values of these metrics depends on the choice of $H$ and $L$ parameters. 
In particular, the value of the minimum number of samples per leaf strongly influences the performance. 
Such parameter implicitly assigns a weight to the classes, thus such values are proportional to the number of occurrences of the class in the dataset. 
In detail, as for Nvidia architecture, \emph{go2} again shows the best performance. By analyzing Figure~\ref{fig:dtpr_p100}, different models (x-axis) achieve high scores (\textbf{DTPR}$> 0.7$) meanwhile the impact of the misclassification of models learned from the other datasets is particular relevant. This result is also evident by analyzing \textbf{DTTR} values in Figure \ref{fig:dttr_p100}. 
For the ARM architecture, the landscape is different. As a matter of fact, overfitted models (see for example the models with $L=0.1$ in Figure~\ref{fig:dtpr_mali}) mitigate the impact of the misclassification improving \textbf{DTPR} scores on \emph{AntonNet}.
On top on the results we showed, \textbf{DTTR} scores also provide a preliminary measure of the performance of the model-driven CLBlast against the standard tuned CLBlast.
From the $DTTR$ results of the models trained from \emph{po2} and \emph{AntonNet} datasets, the model-driven CLBlast library shows the same performance of the traditional tuned version on the Nvidia GPU. 
Likewise, the \textbf{DTPR} scores give a preliminary estimation of how much the models are close to the best possible solution. 
Finally, just for completeness, we report in Table~\ref{Table:go2-P100} all the statistics and metrics for all the decision trees learned from the dataset \emph{go2} on the Nvidia GPU.
By analyzing our metrics, the best model is \textit{hMax-L1} even if \textit{h8-L1} have a higher accuracy ($67$\%). As a consequence, an improvement of the accuracy of the decision trees does not guarantee an improvement in terms of performance.  Regarding Mali GPU, we report in Table~\ref{Table:AntonNet-Firefly} the statistics of the models generated from \emph{AntonNet}.

\subsection{MicroBenchmark}
\begin{figure*}[ht]
  \subfloat[Dataset: \emph{go2}. Model: \emph{hMax-L1}.]
  {
   \includegraphics[scale=0.55]{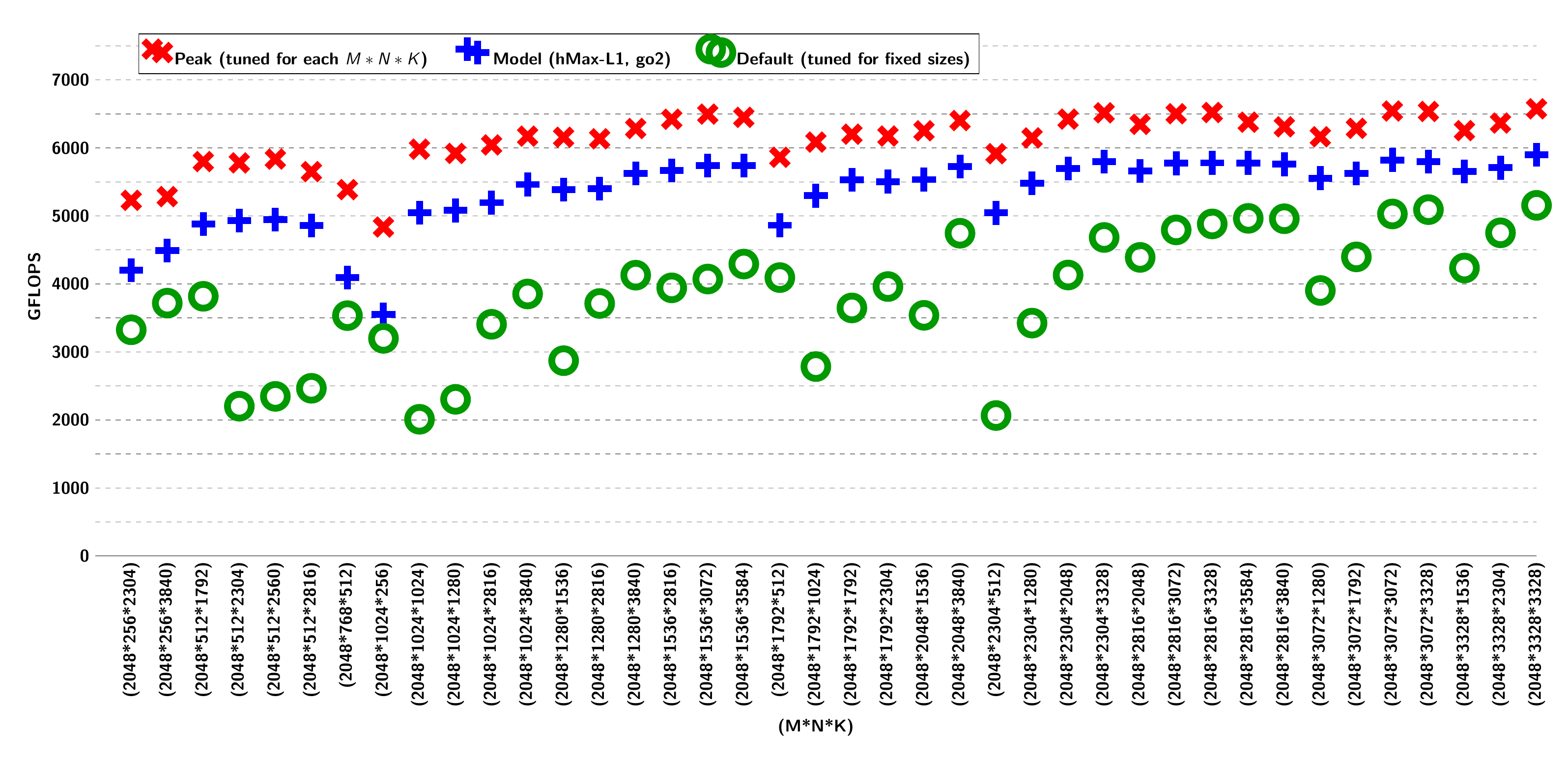}
    \label{fig:go2_p100_2048}
  }
   \quad
  \subfloat[Dataset: \emph{po2}. Model: \emph{hMax-L1}.]
  {
    \includegraphics[scale=0.5]{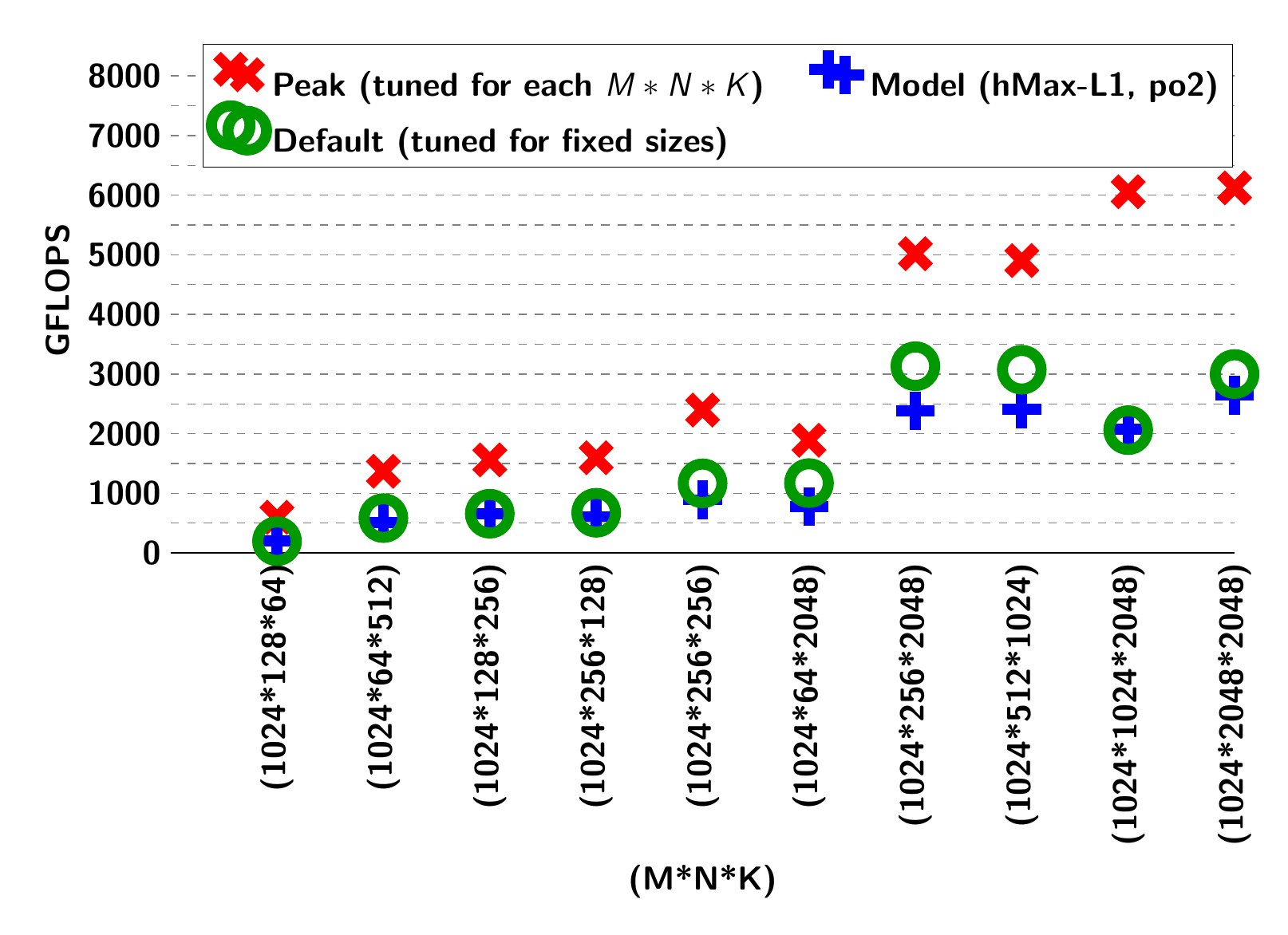}
    \label{fig:po2_p100_1024}
  }
  \subfloat[Dataset: \emph{AntonNet}. Model: \emph{h4-L1}.]
  {
    \includegraphics[scale=0.5]{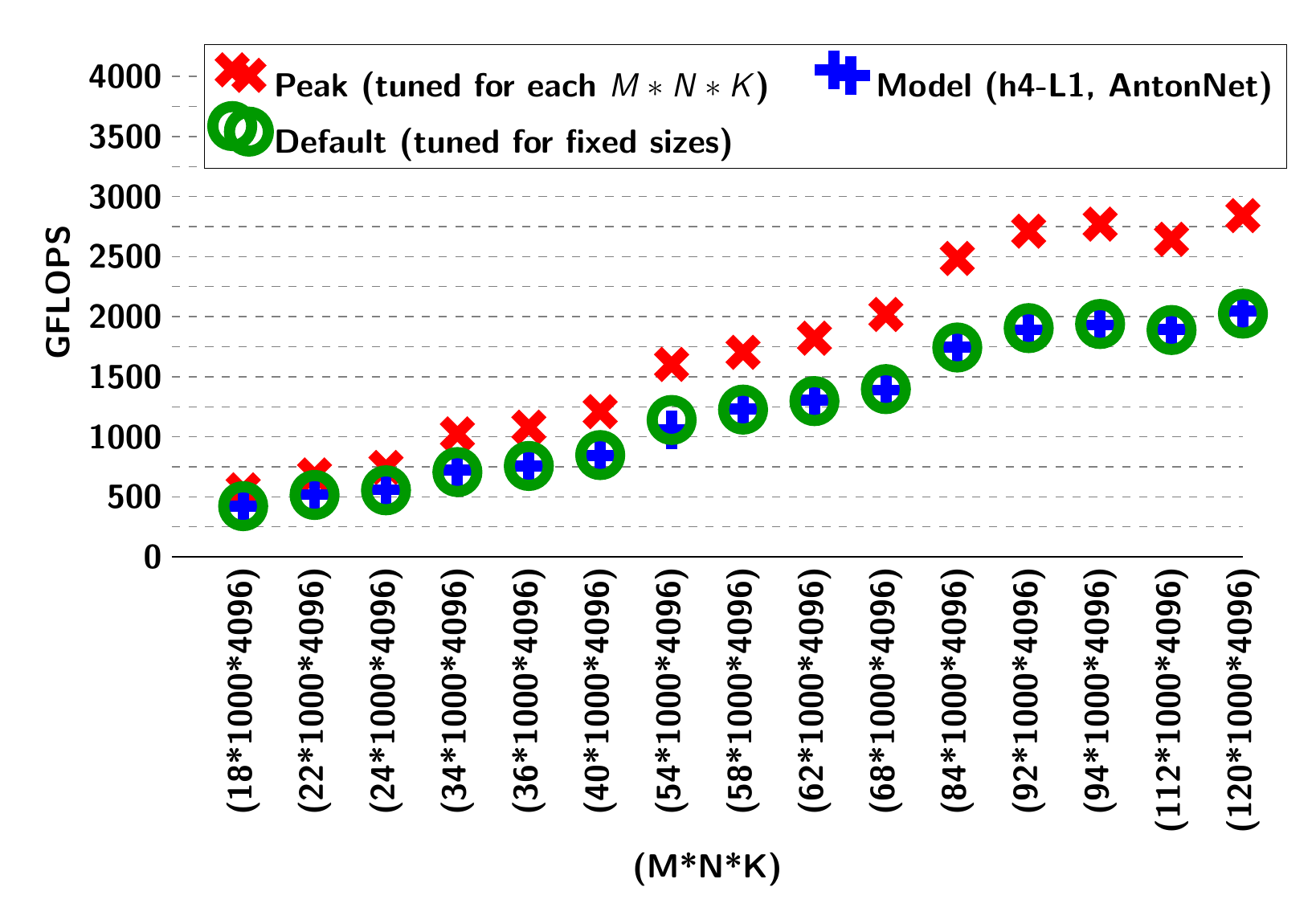}
    \label{fig:AntonNet_p100_1000_4096}
  }
  \caption{Performance evaluation of model-driven CLBlast vs CLBlast traditionally tuned on Nvidia P100.}
  \label{fig:benchmark-nvidia}
\end{figure*}

\begin{figure*}[]
   \subfloat[Dataset: \emph{po2}. Model: \emph{h8-L0.1}.]
   {
     \includegraphics[scale=0.6]{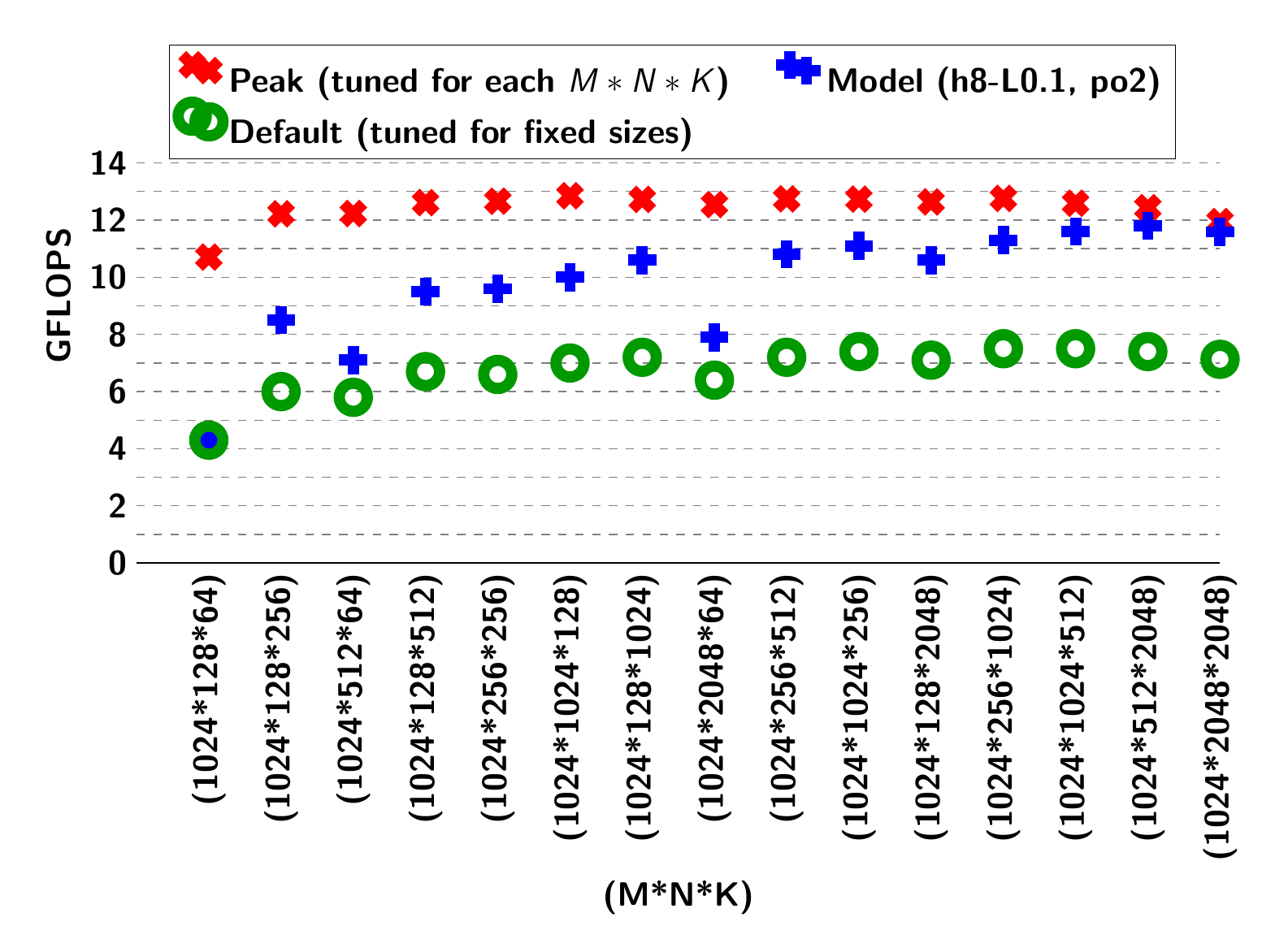}
     \label{fig:po2_mali_1024}
   }
   \subfloat[Dataset: \emph{AntonNet}. Model: \emph{h1-L0.1}.]
   {
     \includegraphics[scale=0.6]{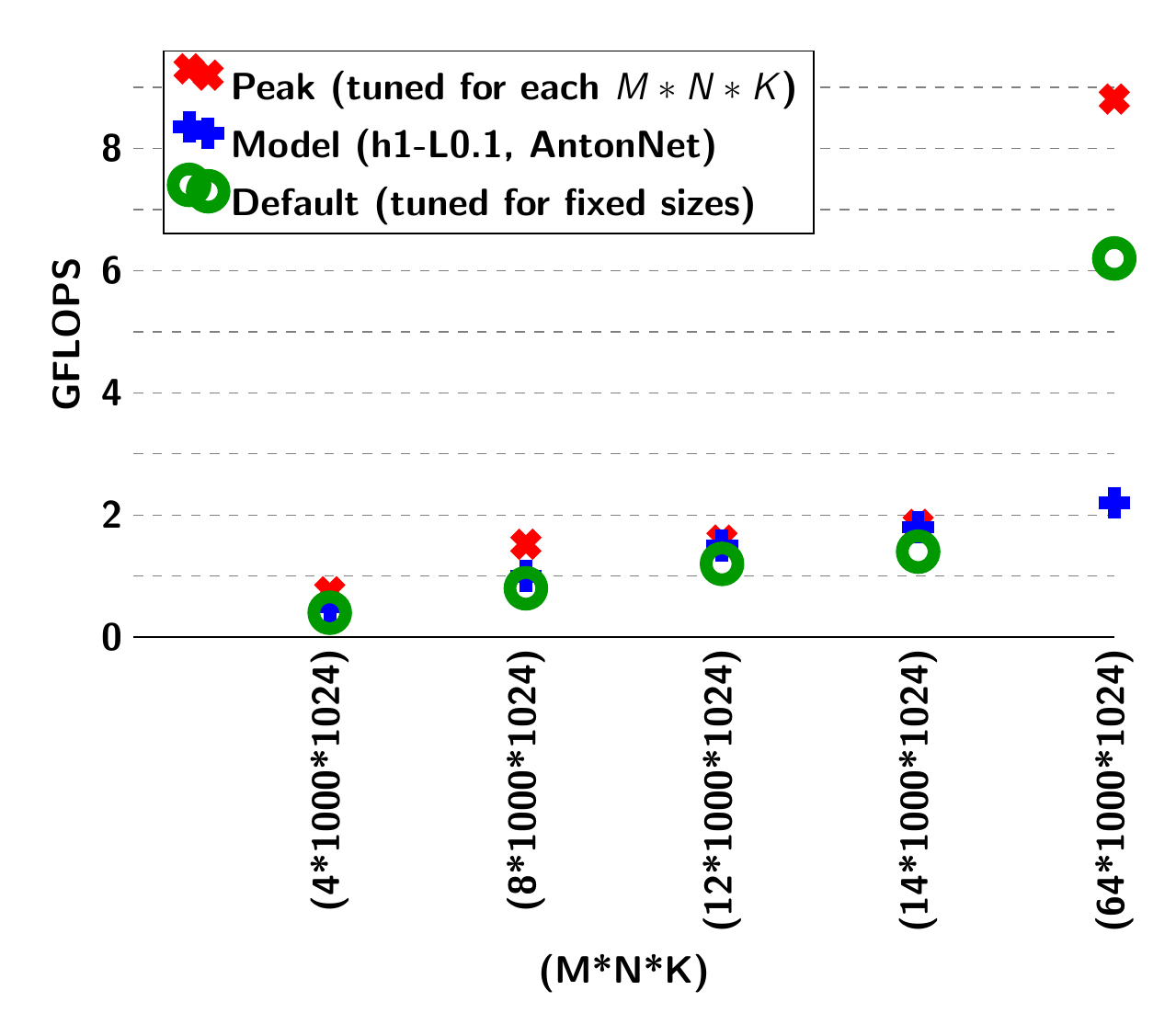}
     \label{fig:AntonNet_mali_1000_1024}
   }
   \caption{Performance evaluation of model-driven CLBlast vs CLBlast traditionally tuned on ARM Mali-T860.}
   \label{fig:benchmark-arm}
 \end{figure*}

The previous experiments showed the average performance ratio of the model-driven CLBLast against both the traditionally tuned CLBlast (v1.0), and the peak performance of the tuner over all the matrices in the test dataset randomly generated. 
For the evaluation of the impact of the misclassification that is important, especially when the goal is evaluating several models that are trained from very specific datasets like \emph{AntonNet}. 
The metrics \textbf{DTPR} and \textbf{DTTR} measure how a model is good in general in terms of performance, overfitting and misclassification. 
Thus, our metrics are good indicators for the selection of the most promising models. 
However, the average of the ratio might not provide a good estimation of the real performance. For example, for specific matrices the improvement may be no relevant when the number of operations (FLOPS) are small (see for example the first triple in Figure~\ref{fig:AntonNet_mali_1000_1024}). 
Thus, hereafter, we show the  performance in GFLOPS, of the model-driven CLBlast against the traditionally tuned CLBlast and the peak performance of the tuner over a wide range of matrices of test datasets. 
In Figure~\ref{fig:benchmark-nvidia} and Figure~\ref{fig:benchmark-arm}, we report the performance of the best model we found for each datasets on Nvidia and ARM architectures respectively.
For Nvidia P100, the model \textit{hMax-L1} learned from \textit{go2} achieves very good performance in most of the points, with the maximum speed-up of 3x over the traditional tuned CLBlast (Figure~\ref{fig:go2_p100_2048}). $DTTR$ shows an improvement of $1.42$x on average. This is mainly due to the modest improvement on the matrices close to the default size used in the traditional tuned version. 
On the contrary, looking at the results of the model learned from a more sparse dataset, the model-driven approach does not guarantee satisfactory performance on average even if in some case it is able to achieve good speed-ups as shown in Figure~\ref{fig:po2_p100_1024}. 
On the Mali-T860, Figure~\ref{fig:po2_mali_1024} surprisingly shows significant speed-ups (up to $2.5$x) for several matrices even if \textbf{DDTR} states a small improvement on the average ($1.12$x). 
For both the architectures, the models learned from \emph{AntonNet} dataset show unsatisfactory performance even if the models have good accuracy on the Mali GPU.  
One reasons is that the decision tree classifier is not able to learn a good model (in term of performance) from a very specific datasets as \emph{AntonNet}. 
Secondly, the misclassification represent an important issue in this case: for the matrices in the dataset, the configurations learned are very specific and different from each other. This means that the wrong configuration (and kernel) selected by the model usually achieves very poor performance.
Third, the gap between CLBlast tuned and the peak of the tuner is not significant (see Figure~\ref{fig:AntonNet_mali_1000_1024}). 
This means that both the tuner and gemm implementations do not achieve good performance \emph{per se}. 
Summarizing about performance, the best models found \textit{hMax-L1} (trained from \emph{go2}) for Nvidia and \textit{h8-L0.1} (trained from \emph{po2}) for ARM outperform the tuned CLBlast as shown in Figure~\ref{fig:go2_p100_2048} and Figure~\ref{fig:po2_mali_1024}. Thus, that models should be used in practice in real applications. 
We finally conclude the section by showing the cost to traverse the decision tree and thus quantify the overhead of the code generated by our framework. 
We analyzed \emph{hMax-L1} model on \emph{go2} (which has $1200$ leaves and depth equal to $19$) over all the matrices in the test dataset. 
The corresponding if-then-else statement introduces less than $2\%$ of overhead on small matrices by selecting the deepest leaf. It definitively decreases as the size of the matrices grows. On average, the overhead impacts less than $1\%$ on performance. We observed a similar trend on the ARM based architecture.

%

\section{Related Work}
\label{sec:related}
There are several papers and notable results that have inspired our works. 
Some of them have been focused on input- and hardware-aware methodologies, meanwhile others target BLAS optimization specifically.
More recently, with the pandemic adoption of the machine learning~\cite{murphy2012machine,alpaydin2014introduction,han2011data}, model-driven approaches come out. 
Auto-tuning and input aware techniques~\cite{falch2015machine, Ding:2015:AAC:2813885.2737969} are recently used to address the problem of performance portability on different data-driven applications~\cite{hou2017auto, Magni:2013:IAD:2458523.2458530, cosenza2017autotuning}. An interesting approach extends such techniques in the presence of multiple algorithmic choice~\cite{7965198}. However, their on-line solution is suitable when a specific routine is called multiple times. 
As for hardware-oblivious approaches, the Nitro framework provides cross-architecture performance portability by building a model on a target architecture from training on different source architectures~\cite{6877283}.
Specifically to BLAS, several optimized linear algebra and BLAS libraries have been released~\cite{xianyi2014openblas, whaley1998automatically, anderson1990lapack, choi1992scalapack}.
Some of them have been designed for accelerators~\cite{CLBlast2017, du2012cuda, rupp2010viennacl} or for specific GPU architectures only~\cite{nvidia2008cublas}.
Several works have previously published auto-tuning and optimization approaches to accelerate GEMM~\cite{lai2013performance, li2009note, matsumoto2012performance}. 
The problem of the exploration of huge search space of tunable parameters has been partially mitigated by the use of meta-heuristics optimization approaches ~\cite{CLTune2015,Werkhoven2014} and machine learning techniques~\cite{Tillet:2017:IAC:3126908.3126939, falch2015machine}. The formers are able to predict parameters by starting from the exploration of a small search space~\cite{6339587, 7967118, CPE:CPE4029}.
From industrial prospective, vendors libraries (e.g., MKL, cuBLAS and ARM Compute Library) still apply manual heuristics in order to select at runtime highly-optimized code for specific inputs. 
Contrarily to those solutions, recently model-driven solutions have been adopted for selecting the best numerical method to solve the linear advection equation~\cite{arteaga2017model} and optimizing sparse CP decomposition~\cite{li2017model}. Others investigated machine learning techniques to accelerate sparse linear algebra operations~\cite{Choi:2010:MAS:1693453.1693471, elafrou2017performance, zhao2017poster}. 
Tillet et al., developed ISAAC, which exploits a multi-layer perceptron (MLP) to generate high optimized parametric-code in the training step, such that at run-time, the library infers the best parameters for the specific input~\cite{Tillet:2017:IAC:3126908.3126939}. However, since it generates assembly code, it is not able to run on different architectures like ARM. 
Contrarily to the existing works, our solution is general since it can be applied to different architectures and problems. 
Especially for the architecture perspective, we do not need the exposure of the instruction set as ISAAC~\cite{Tillet:2017:IAC:3126908.3126939} requires. 
This also makes our solution robust since it is not affected by architectural changes.

\section{Conclusion}
\label{sec:conclusion}

When designing high performance applications, a key problem is how to select the optimal algorithm/implementation/configuration for a given combination of data types, data sizes, system capabilities, etc.
In this paper, we presented a machine learning based approach to building highly-optimized adaptive libraries for data-driven applications.
We analyzed a simple white-box supervised classifier to build a predictive model for GEMM on GPUs.
We analyzed in depth the performance of several models trained from different datasets and generated by tuning different parameters. While decision trees did not achieve particularly high accuracy, we still observed significant performance improvements (up to 3x) compared to the traditional, non-adaptive approach: in practice, the impact of mispredictions is mitigated when the model is generated from a dense dataset, even when using just a few features. 
We validated this approach with a production-quality BLAS OpenCL library on two very different GPU architectures. We are planning to release our source code and the datasets as customizable and reusable Collective Knowledge components.

We are extending this work in several directions. First, we are investigating advanced ML techniques to generate more effective models, especially when the training datasets are small and potentially specific (like AntonNet). Second, we looking into how to generate more compact but still representative training sets. This aspect is particularly crucial for embedded architectures where generating the training set is expensive (e.g., it took 7 days to create \emph{po2} for the Mali GPU). We believe in a collaborative/community-driven approach for collecting and analyzing datasets, building predictive models, etc.~\cite{7459430}.

Finally, we are studying more complex problems such as graph analytics, where it is hard to predict the computation due to many possible choices for data structures (e.g. CSR or COO)~\cite{zhao2017poster}, data-thread mapping strategies (vertex or edge parallelism) and algorithms (e.g. top-down or bottom-up~\cite{8267334}).

\section{Acknowledgements}
Marco Cianfriglia was supported by HiPEAC projec ``Industrial PhD Internship 2016''.
The HiPEAC project has received funding from the European Union’s Horizon 2020 research and innovation programme under grant agreement number 779656.

\bibliographystyle{abbrv}

\input{revised.bbl}
\input{table_go2_p100}

\input{table_antonNet_mali}

\end{document}

%% file: table_hardware.tex
\begin{table}[ht]
\centering
\scriptsize

\begin{tabular}{|c|c|c|}
\hline
\textbf{Device name} &\textbf{Nvidia P100} & \textbf{ARM Mali-T860}\\ \hline
\textbf{Market segment} & Server & System on Chip \\ \hline
\textbf{Micro-architecture} & Pascal & Midgard 4th gen\\ \hline
\textbf{Number of available cores } & 3584 CUDA cores  & 4 Mali cores \\
& (GP100) & \\ \hline 
\textbf{Boost frequency} & 1353 MHz &  2000 MHz \\ \hline
\textbf{Processing power} & 9.7 TFLOPS & 23.8 GFLOPS \\ \hline
\textbf{Memory available} & 16 GB & 4 GB \\ \hline
\textbf{Memory type} &HBM2 & DDR3 \\ \hline
\end{tabular}

\caption{Nvidia P100 and ARM Mali-T860 hardware description.}
\label{table:hardware}
\end{table}

%% file: table_summary.tex
\begin{table*}
\centering
\scriptsize

\begin{tabular}{|c|c|c|c|c|c|c|c|}
\hline
    \textbf{Dataset}           &  \textbf{Dataset}   & \textbf{Number of}  & \textbf{Number of} & \textbf{Best} & \textbf{Best} & \textbf{Best} & \textbf{Best} \\
    \textbf{Name} & \textbf{Size}  & \textbf{Unique Config.} & \textbf{Unique Config.}  & \textbf{Decision Tree}& \textbf{Decision Tree}& \textbf{Decision Tree}& \textbf{Decision Tree}\\
    & & \textbf{Xgemm} & \textbf{XgemmDirect} & \textbf{Name} & \textbf{accuracy} & \textbf{DTPR} & \textbf{DTTR}\\ 
    \hline
    AntonNet & 456 & 1 & 81& h4-L1 & 36 & 0.484 & 1.013   \\ \hline
    PowerOf2(po2) & 216 & 2 & 41 &  hMax-L1 & 21  & 0.431  & 0.931\\ \hline
    GridOf2(go2) & 3375 & 6 & 22 & hMax-L1 & 60 & 0.852 & 1.424\\ \hline
\end{tabular}

\caption{Datasets statistics - Nvidia P100. ``Best Decision Tree'' refers to the model with the highest \textbf{DTPR} score. The sum (per row) of the columns 3 and 4 represents the total number of classes of the dataset.}
\label{table:stats-p100}
\end{table*}

\begin{table*}
\centering
\scriptsize

\begin{tabular}{|c|c|c|c|c|c|c|c|}
\hline
    \textbf{Dataset}           &  \textbf{Dataset}   & \textbf{Number of}  & \textbf{Number of} & \textbf{Best} & \textbf{Best} & \textbf{Best} & \textbf{Best} \\
    \textbf{Name} & \textbf{Size}  & \textbf{Unique Config.} & \textbf{Unique Config.}  & \textbf{Decision Tree}& \textbf{Decision Tree}& \textbf{Decision Tree}& \textbf{Decision Tree}\\
    & & \textbf{Xgemm} & \textbf{XgemmDirect} & \textbf{Name} & \textbf{accuracy} & \textbf{DTPR} & \textbf{DTTR}\\ 
    \hline
    AntonNet & 456 & 28 & 35&  h1-L0.1 & 55 & 0.702 & 1.092  \\ \hline
    PowerOf2(po2) & 216 & 29 & 1 &  h8-L0.1 & 45 & 0.551 & 1.121\\ \hline

\end{tabular}
\caption{Dataset statistics - ARM Mali-T860. ``Best Decision Tree'' refers to the model with the highest \textbf{DTPR} score. The sum (per row) of the columns 3 and 4 represents the total number of classes of the dataset.}
\label{table:stats-mali}
\end{table*}

%% file: table_go2_p100.tex
\begin{table*}
\scriptsize
\centering
  
  \begin{tabular}{|c|c|c|c|c|c|c|c|c|c|c|}
    \hline
    \textbf{Decision Tree}   & \textbf{Accuracy} & \textbf{DTPR}  & \textbf{DTTR}  & \textbf{Total} & \textbf{Decision Tree} & \textbf{Min}  & \textbf{Number of} & \textbf{Number of} & \textbf{Number of} & \textbf{Number of}  \\
    \textbf{Name} &\textbf{(\%)} & & &\textbf{number of} &\textbf{Height} & \textbf{Samples} & \textbf{Unique Config.} & \textbf{Unique Config.} & \textbf{Leaves} & \textbf{Leaves}\\ 
    &  & & &\textbf{Leaves} & & \textbf{PerLeaf}& \textbf{Gemm} & \textbf{GemmDirect} &\textbf{Gemm} & \textbf{GemmDir}\\ \hline
    h1-L1     & 62       & 0.376 & 0.637 & 2           & 1         & 1      & 1 & 1       & 1 & 1      \\ \hline
    h1-L2     & 62       & 0.376 & 0.637 & 2           & 1         & 2        & 1 & 1     & 1 & 1     \\ \hline
    h1-L4     & 62       & 0.376 & 0.637 & 2           & 1         & 4           & 1 & 1    & 1 & 1   \\ \hline
    h1-L0.1   & 62       & 0.376 & 0.637 & 2           & 1         & 0.1        & 1 & 1    & 1 & 1    \\ \hline
    h1-L0.2   & 62       & 0.376 & 0.637 & 2           & 1         & 0.2        & 1 & 1   & 1 & 1     \\ \hline
    h1-L0.3   & 59       & 0.436 & 0.736 & 2           & 1         & 0.3       & 0 & 2  & 0 & 2       \\ \hline
    h1-L0.4   & 56       & 0.444 & 0.735 & 2           & 1         & 0.4       & 1 & 1     & 1 & 1    \\ \hline
    h1-L0.5   & 51.5     & 0.433 & 0.734 & 1           & 0         & 0.5        & 0 & 1    & 0 & 1    \\ \hline
    h2-L1     & 62       & 0.433 & 0.734 & 4           & 2         & 1         & 1 & 2   & 2 &  2    \\ \hline
    h2-L2     & 62       & 0.416 & 0.703 & 4           & 2         & 2         & 1 & 2      & 2 &  2   \\ \hline
    h2-L4     & 62       & 0.415 & 0.702 & 4           & 2         & 4          & 1 & 2     & 2 &  2   \\ \hline
    h2-L0.1   & 62       & 0.415 & 0.702 & 4           & 2         & 0.1         & 1 & 2    & 2 &  2   \\ \hline
    h2-L0.2   & 62       & 0.416 & 0.703 & 3           & 2         & 0.2        & 1 & 2   & 1 &  2     \\ \hline
    h2-L0.3   & 59       & 0.416 & 0.982 & 3           & 2         & 0.3      & 0 & 3      & 0 &  3    \\ \hline
    h2-L0.4   & 56       & 0.606 & 0.736 & 2           & 1         & 0.4         & 1 & 1    & 1 &  1   \\ \hline
    h2-L0.5   & 51.5     & 0.445 & 0.734 & 1           & 0         & 0.5        & 0 & 1      & 0 &  1  \\ \hline
    h4-L1     & 67       & 0.687 & 1.120 & 16          & 4         & 1        & 1 & 5     & 2 &  14     \\ \hline
    h4-L2     & 67       & 0.688 & 1.122 & 16          & 4         & 2           & 1 & 5    & 2 &  14    \\ \hline
    h4-L1     & 67       & 0.686 & 1.119 & 16          & 4         & 4          & 1 & 5      & 2 &  14   \\ \hline
    h4-L0.1   & 65.5     & 0.576 & 0.931 & 8           & 4         & 0.1       & 1 & 4     & 2 &  6     \\ \hline
    h4-L0.2   & 62       & 0.506 & 0.845 & 4           & 3         & 0.2      & 1 & 3      & 1 &  3     \\ \hline
    h4-L0.3   & 59       & 0.605 & 0.981 & 3           & 2         & 0.3       & 0 & 3     & 0 &  3     \\ \hline
    h4-L0.4   & 56       & 0.445 & 0.737 & 2           & 1         & 0.4       & 1 & 1      & 1 &  1    \\ \hline
    h4-L0.5   & 51.5     & 0.434 & 0.735 & 1           & 0         & 0.5       & 0 & 1       &0 &  1   \\ \hline
    h8-L1     & 67       & 0.806 & 1.340 & 215         & 8         & 1        & 1 & 9     & 4 & 211       \\ \hline
    h8-L2     & 66.5     & 0.807 & 1.341 & 201         & 8         & 2                & 1 & 8  & 4 &  197 \\ \hline
    h8-L4     & 66       & 0.806 & 1.304 & 175         & 8         & 4        & 1 & 6       & 4 &  171    \\ \hline
    h8-L0.1   & 65.5     & 0.576 & 0.931 & 8           & 4         & 0.1      & 1 & 4      & 2& 6    \\ \hline
    h8-L0.2   & 62       & 0.506 & 0.845 & 4           & 3         & 0.2      & 1 & 3    & 1 &3      \\ \hline
    h8-L0.3   & 59       & 0.606 & 0.982 & 3           & 2         & 0.3       & 0 & 3   &0&3      \\ \hline
    h8-L0.4   & 56       & 0.445 & 0.736 & 2           & 1         & 0.4        & 1 & 1    &1 &1    \\ \hline
    h8-L0.5   & 51.5     & 0.433 & 0.734 & 1           & 0         & 0.5       & 0 & 1    &0&1     \\ \hline
   \textbf{hMax-L1}   & \textbf{60}       & \textbf{0.852} & \textbf{1.424} & \textbf{1290}        & \textbf{19} & \textbf{1}         & \textbf{1} & \textbf{11}    &\textbf{4} & \textbf{1286}     \\ \hline
    hMax-L2   & 58.5     & 0.848 & 1.418 & 790         & 18        & 2          & 1 & 8    &4 & 786    \\ \hline
    hMax-L4   & 64       & 0.846 & 1.412 & 430         & 15        & 4         & 1 & 6    &4 & 426     \\ \hline
    hMax-L0.1 & 65.5     & 0.574 & 0.927 & 8           & 4         & 0.1       & 1 & 4   & 2 & 6      \\ \hline
    hMax-L0.2 & 62       & 0.506 & 0.844 & 4           & 3         & 0.2      & 1 & 3    & 1 & 3      \\ \hline
    hMax-L0.3 & 59       & 0.606 & 0.982 & 3           & 2         & 0.4      & 0 & 3   & 0 & 3       \\ \hline
    hMax-L0.4 & 56       & 0.445 & 0.737 & 2           & 1         & 0.4       & 1 & 1  & 1 & 1       \\ \hline
    hMax-L0.5 & 51.5     & 0.433 & 0.734 & 1           & 0         & 0.5     & 0 & 1     & 0 & 1      \\ \hline
    \end{tabular}
    \caption{Statistics of the decision trees trained from \emph{go2} dataset by varying $H$ and $L$ on the Nvidia P100. The model with the highest \textbf{DTPR} score is reported in bold.}
    \label{Table:go2-P100}
\end{table*}

%% file: table_antonNet_mali.tex
\begin{table*}
\scriptsize
\centering
  \begin{tabular}{|c|c|c|c|c|c|c|c|c|c|c|}
    \hline
    \textbf{Decision Tree}   & \textbf{Accuracy} & \textbf{DTPR}  & \textbf{DTTR}  & \textbf{Total} & \textbf{Decision Tree} & \textbf{Min}  & \textbf{Number of} & \textbf{Number of} & \textbf{Number of} & \textbf{Number of}  \\
    \textbf{Name} &\textbf{(\%)} & & &\textbf{number of} &\textbf{Height} & \textbf{Samples} & \textbf{Unique Config.} & \textbf{Unique Config.} & \textbf{Leaves} & \textbf{Leaves}\\ 
    &  & & &\textbf{Leaves} & & \textbf{PerLeaf}& \textbf{Gemm} & \textbf{GemmDirect} &\textbf{Gemm} & \textbf{GemmDir}\\ \hline
   h1-L1     & 55            & 0.692 & 1.085 & 2           & 1         & 1      &0 &2      &0 &2       \\ \hline
    h1-L2     & 55            & 0.560 & 0.828 & 2           & 1         & 2        &0 &2    &0 &2      \\ \hline
    h1-L4     & 55            & 0.600 & 0.895 & 2           & 1         & 4        &0 &2    &0 &2      \\ \hline
    \textbf{h1-L0.1}   & \textbf{55}            & \textbf{0.702} & \textbf{1.092} & \textbf{2}           & \textbf{1}         & \textbf{0.1}       &\textbf{0} &\textbf{2}   &\textbf{0} &\textbf{2}      \\ \hline
    h1-L0.2   & 55            & 0.631 & 0.955 & 2           & 1         & 0.2       &0 &2     &0 &2    \\ \hline
    h1-L0.3   & 42            & 0.619 & 0.918 & 2           & 1         & 0.3         &0 &1   &0 &2    \\ \hline
    h1-L0.4   & 42            & 0.559 & 0.822 & 2           & 1         & 0.4      &0 &1    &0 &2      \\ \hline
    h1-L0.5   & 42            & 0.418 & 0.691 & 2           & 1         & 0.5      &0 &2   &0 &2       \\ \hline
    h2-L1     & 52.5          & 0.638 & 1.012 & 4           & 2         & 1          &0 &2    &0 &4    \\ \hline
    h2-L2     & 52.5          & 0.544 & 0.823 & 4           & 2         & 2           &0 &2  &0 &4     \\ \hline
    h2-L4     & 52.5          & 0.500 & 0.749 & 4           & 2         & 4         &0 &2    &0 &4     \\ \hline
    h2-L0.1   & 55            & 0.572 & 0.863 & 4           & 2         & 0.1        &0 &2  &0 &4      \\ \hline
    h2-L0.2   & 55            & 0.540 & 0.820 & 3           & 2         & 0.2      &0 &2   &0 &3       \\ \hline
    h2-L0.3   & 42            & 0.555 & 0.831 & 2           & 1         & 0.3       &0 &1   &0 &2      \\ \hline
    h2-L0.4   & 42            & 0.560 & 0.838 & 2           & 1         & 0.4        &0 &1   &0 &2     \\ \hline
    h2-L0.5   & 42            & 0.499 & 0.715 & 2           & 1         & 0.5        &0 &2   &0 &2     \\ \hline
    h4-L1     & 56.5          & 0.641 & 1.005 & 16          & 4         & 1          &1 &2   &1 & 15     \\ \hline
    h4-L2     & 58            & 0.517 & 0.781 & 16          & 4         & 2          &1 &2      &0 &15  \\ \hline
    h4-L1     & 56.5          & 0.677 & 1.062 & 15          & 4         & 4            &1 &2   &0 &14   \\ \hline
    h4-L0.1   & 55            & 0.577 & 0.878 & 7           & 4         & 0.1         &0 &4    &0 &7   \\ \hline
    h4-L0.2   & 55            & 0.446 & 0.681 & 4           & 3         & 0.2          &0 &3  &0 &4    \\ \hline
    h4-L0.3   & 42            & 0.502 & 0.742 & 2           & 1         & 0.3            &0 &1  &0 &2  \\ \hline
    h4-L0.4   & 42            & 0.529 & 0.778 & 2           & 1         & 0.4       &0 &1    &0 &2     \\ \hline
    h4-L0.5   & 42            & 0.440 & 0.617 & 2           & 1         & 0.5        &0 &2   &0 &2     \\ \hline
    h8-L1     & 55            & 0.584 & 0.863 & 84          & 8         & 1       &5 &13    &6 &78       \\ \hline
    h8-L2     & 56.5          & 0.466 & 0.669 & 60          & 8         & 2        &2 &9   &3 &57      \\ \hline
    h8-L4     & 52.5          & 0.551 & 0.826 & 45          & 8         & 4          &1 &7   &2 &43     \\ \hline
    h8-L0.1   & 55            & 0.473 & 0.682 & 8           & 5         & 0.1        &0 &5   &0 &8     \\ \hline
    h8-L0.2   & 55            & 0.466 & 0.669 & 4           & 3         & 0.2      &0 &3   &0 &4     \\ \hline
    h8-L0.3   & 42            & 0.571 & 0.850 & 2           & 1         & 0.3          &0 &1   &0 &2   \\ \hline
    h8-L0.4   & 42            & 0.592 & 0.885 & 2           & 1         & 0.4       &0 &1    &0 &2     \\ \hline
    h8-L0.5   & 42            & 0.591 & 0.865 & 2           & 1         & 0.5        &0 &2    &0 &2    \\ \hline
    hMax-L1   & 52.5          & 0.846 & 1.008 & 166         & 17        & 1     & 9 & 16     &15 &151        \\ \hline
    hMax-L2   & 54            & 0.570 & 0.858 & 95          & 15        & 2       & 3 & 12      &4 &91     \\ \hline
    hMax-L4   & 52.5          & 0.554 & 0.815 & 53          & 10        & 4          &1 & 7   &2 &51     \\ \hline
    hMax-L0.1 & 55            & 0.487 & 0.708 & 8           & 5         & 0.1        &0 &5   &0 &8     \\ \hline
    hMax-L0.2 & 55            & 0.438 & 0.667 & 4           & 3         & 0.2         &0 &3   &0 &4      \\ \hline
    hMax-L0.3 & 42            & 0.628 & 0.954 & 2           & 1         & 0.3            &0 &1  &0 &2    \\ \hline
    hMax-L0.4 & 42            & 0.604 & 0.895 & 2           & 1         & 0.4       &0 &1    &0 &2       \\ \hline
    hMax-L0.5 & 42            & 0.496 & 0.714 & 2           & 1         & 0.5         &0 &2    &0 &2     \\ \hline
    \end{tabular}
    
    \caption{Statistics of the decision trees trained from \emph{AntonNet} dataset by varying $H$ and $L$ on the ARM Mali-T860. The model with the highest \textbf{DTPR} score is reported in bold.}
    \label{Table:AntonNet-Firefly}
\end{table*}